\documentclass[12pt,nofootinbib,preprint,superscriptaddress]{revtex4}
\usepackage{epsfig,subfigure}
\usepackage{epstopdf}
\usepackage[utf8]{inputenc}
\usepackage{graphicx}
\usepackage{amssymb}
\usepackage{enumitem}
\usepackage{amsxtra}
\usepackage{amsmath}
\usepackage[export]{adjustbox}
\usepackage{booktabs,multirow,tabularx}
\usepackage{slashed}
\usepackage{float}
\usepackage{physics}
\usepackage{placeins}
\usepackage{rotating}
\usepackage{lscape}
\usepackage{color}
\usepackage{hyperref}
\usepackage{soul}
\usepackage{bm}

\newcommand{\bea}{\begin{eqnarray}}
\newcommand{\eea}{\end{eqnarray}}

\def\beq{\begin{equation}}
\def\eeq{\end{equation}}

\usepackage{siunitx}
\usepackage{stackengine}
\newcommand\vew{v_{\rm EW}}

\begin{document}

\title{Real singlet scalar benchmarks in the multi-TeV resonance regime}

\author{Ian M. Lewis}
\email{ian.lewis@ku.edu}
\affiliation{Department of Physics and Astronomy, University of Kansas, Lawrence, Kansas, 66045~ USA}

\author{Jacob L. Scott}
\email{jacob.scott@ku.edu}
\affiliation{Department of Physics and Astronomy, University of Kansas, Lawrence, Kansas, 66045~ USA}

\author{Miguel A. Soto Alcaraz}
\email{msoto1@ku.edu}
\affiliation{Department of Physics and Astronomy, University of Kansas, Lawrence, Kansas, 66045~ USA}

\author{Matthew Sullivan}
\email{msullivan1@bnl.gov}
\affiliation{High Energy Theory Group, Physics Department, Brookhaven National Laboratory,
Upton, New York, 11973~ USA}

\begin{abstract}
Scalar extensions of the Standard Model are of much interest at the LHC and future colliders.  In particular, these models can give rise to resonant di-Higgs production and alter the Higgs trilinear coupling.  In this paper, we study di-Higgs production in the Standard Model extended by a real scalar singlet with no additional symmetries.  We determine how large the resonant di-Higgs rate and variation in the Higgs trilinear coupling can be in four scenarios: current LHC results and projected results at the HL-LHC, the HL-LHC combined with a circular $e^-e^+$ collider such as the CEPC or FCC-ee, and the HL-LHC combined with a linear $e^-e^+$ collider such as the ILC.  While these are updated results from a previous study~\cite{Lewis:2017dme} using current LHC data, we go further and find benchmark points in the multi-TeV resonance regime for future colliders beyond the HL-LHC. Considering current LHC results, the resonant di-Higgs rate can still be an order of magnitude larger than the SM predicted di-Higgs rate.  In the HL-LHC scenario, the Higgs trilinear coupling can still be a factor of three larger than the SM prediction for resonance masses in the $1.5-3.5$ TeV range, where resonant searches may have less reach. This enhancement is just at the projected 2$\sigma$ sensitivity of the HL-LHC.  We find there are resonance masses for which the change in the Higgs trilinear is maximized while the resonant rate is negligible.  We provide an analytical understanding of these effects with a discussion on the interplay of various constraints on the parameter space and the Higgs trilinear coupling. 
\end{abstract}

\maketitle



\section{Introduction}
\label{introduction}
Measurements of the observed Higgs boson properties are some of the major long term goals of the Large Hadron Collider (LHC) and future colliders~\cite{Dawson:2022zbb,EuropeanStrategyforParticlePhysicsPreparatoryGroup:2019qin}.  Of particular interest is the measurement of the Higgs potential~\cite{DiMicco:2019ngk}, which is the source of electroweak symmetry breaking (EWSB) within the Standard Model (SM).  So far only the quadratic term in the potential has been directly measured via the Higgs mass. The most promising route to further directly measure the shape of the potential is to measure the cubic term via the Higgs trilinear coupling and di-Higgs production.  The LHC is projected to limit the trilinear coupling to be between zero and 2.7 times the SM value at the $2\sigma$ level~\cite{ATL-PHYS-PUB-2022-005}.  Hence, models that alter the di-Higgs rate are of much interest~\cite{DiMicco:2019ngk}.

In this paper we study di-Higgs production in the simplest extension of the SM: the addition of a real gauge singlet scalar with no additional symmetries.  This model and signal has been much studied in the literature with and without an additional $Z_2$ symmetry on the new scalar~\cite{Silveira:1985rk,OConnell:2006rsp,Barger:2007im,Bowen:2007ia,
No:2013wsa,Pruna:2013bma,Profumo:2014opa,Chen:2014ask,Dawson:2015haa,Buttazzo:2015bka,Robens:2015gla,
Robens:2016xkb,Lewis:2017dme,DiLuzio:2017tfn,Huang:2017jws,Dawson:2018dcd,Li:2019tfd,Alves:2020bpi,Dawson:2020oco,Adhikari:2020vqo,Cohen:2020xca,Dawson:2021jcl,Hammad:2023sbd,Feuerstake:2024uxs,Goncalves:2024vkj,Profumo:2007wc,Kotwal:2016tex,Papaefstathiou:2020iag,Papaefstathiou:2021glr}. Without the $Z_2$ symmetry, there are more free parameters with which to increase the trilinear couplings. The real singlet scalar only couples to the SM via the scalar potential at the renormalizable level.\footnote{Nonrenormalizable interaction can introduce new couplings
between the singlet scalar and gauge bosons and fermions that affect interpretations of the simplest model~\cite{Adhikari:2020vqo}.} Since the shape of the Higgs potential is changed, this model can help provide a strong first order electroweak (EW) phase transition necessary for EW baryogenesis~\cite{Profumo:2007wc,Espinosa:2011ax,Curtin:2014jma,Kotwal:2016tex,Vaskonen:2016yiu,Huang:2016cjm,Beniwal:2017eik,Kurup:2017dzf,Chen:2017qcz,Beniwal:2018hyi,Ghorbani:2018yfr,Mazumdar:2018dfl,Papaefstathiou:2020iag,Ghorbani:2020xqv,Papaefstathiou:2021glr}.  That is, it is a well-motivated model that introduces interesting new physics related to the Higgs sector.  Hence, the real singlet scalar model is a good benchmark model for the LHC and future colliders.

The main results of this paper are finding the maximum allowed resonant di-Higgs rates and variation in the Higgs trilinear coupling in this model.  This is an update on previous results~\cite{Lewis:2017dme} to take into account current LHC bounds~\cite{Lane:2024vur}, as well as extending that work to provide benchmarks assuming various future collider possibilities. In addition to the high luminosity LHC (HL-LHC), we consider several different scenarios with precision electron-positron colliders.  For example, there are the possibilities of linear electron-positron colliders such as the International Linear Collider (ILC)~\cite{Behnke:2013xla,ILC:2013jhg,Adolphsen:2013jya}, the Circular Electron Positron Collider (CEPC)~\cite{CEPCStudyGroup:2018rmc,CEPCStudyGroup:2018ghi} or the Future Circular Collider with electron-positron collisions (FCC-ee)~\cite{FCC:2018evy}. We consider four scenarios~\cite{Lane:2024vur}:
\begin{enumerate}[label=(\roman*)]
\item S1:~Current LHC constraints from precision Higgs measurements and scalar searches~\cite{Lane:2024vur}.
\item S2:~Projections for Higgs precision measurements and scalar searches at the HL-LHC~\cite{EuropeanStrategyforParticlePhysicsPreparatoryGroup:2019qin,Dawson:2022zbb}.
\item S3:~Projections for Higgs precision measurements and scalar searches at the HL-LHC and the CEPC or FCC-ee (HL-LHC+FCCee)~\cite{EuropeanStrategyforParticlePhysicsPreparatoryGroup:2019qin,Dawson:2022zbb}
\item S4:~Projections for Higgs precision measurements and scalar searches at the HL-LHC and a 500 GeV ILC (HL-LHC+ILC500)~\cite{EuropeanStrategyforParticlePhysicsPreparatoryGroup:2019qin,Dawson:2022zbb}.
\end{enumerate}
Scenario S1 is the update on Ref.~\cite{Lewis:2017dme} and provides benchmark points for the LHC and the HL-LHC. Since these are benchmark points for the LHC, resonant masses up to 1 TeV are considered where HL-LHC direct searches are expected to be most constraining~\cite{EuropeanStrategyforParticlePhysicsPreparatoryGroup:2019qin,Dawson:2022zbb}.   Scenarios S2-S4 project benchmark points for future high energy colliders up to resonance masses of 10 TeV.  The results for scenarios S2-S4 are presented in such a way as to be relevant for most future high energy collider possibilities, including the HL-LHC, 100 TeV pp colliders~\cite{Tang:2015qga,CEPC-SPPCStudyGroup:2015csa,FCC:2018vvp} or a future muon collider~\cite{Palmer:1995jy,Delahaye:2019omf,Black:2022cth,MuonCollider:2022xlm}.
We find that there can still be significant enhancements in the Higgs trilinear coupling in the TeV to multi-TeV range, up to a factor of three larger than the SM predictions.  Such resonances may be out of the reach of the LHC, and a variation in the Higgs trilinear may be the first indication of new physics.

In Sec.~\ref{Model} we provide a description of the model.  Various theoretical and experimental constraints are presented in Sec.~\ref{constraints}.  The results for the maximum di-Higgs rate as well as the allowed ranges for the Higgs trilinear couplings in scenarios S1-S4 are given in Sec.~\ref{sec:Results}.  We conclude in Sec.~\ref{sec:conc}.

\section{Higgs Singlet Model}
\label{Model}
The model extends the SM by adding a real gauge singlet scalar $S=x+s$, where $\langle S \rangle = x$ is the vacuum expectation value (vev) of $S$.  This new singlet scalar can only couple directly to the SM at the renormalizable level via the scalar potential~\cite{Chen:2014ask}. 
\begin{equation}
    \begin{aligned}
    V(\Phi,S) = & -\mu^2 \Phi^{\dag}\Phi + \lambda\left(\Phi^{\dag}\Phi\right)^2 + \frac{a_1}{2}\Phi^{\dag}\Phi S + \frac{a_2}{2}\Phi^{\dag}\Phi S^2  \\[10pt]
    & + b_1S + \frac{b_2}{2}S^2 + \frac{b_3}{3}S^3 + \frac{b_4}{4}S^4,
    \end{aligned}
    \label{eq:Potential}
\end{equation}
where 
\begin{eqnarray}
\Phi=\frac{1}{\sqrt{2}}\begin{pmatrix} \sqrt{2}G^+ \\ v+h+i\,G^0\end{pmatrix}
\end{eqnarray}
is the Higgs doublet, $v$ is the SM Higgs vev, $h$ is the SM Higgs boson, and $G^+,G^0$ are the Goldstone bosons.

Shifting the field $S$ by a constant value $S\rightarrow S+\delta S$ is simply a redefinition of parameters with no physical consequence~\cite{Chen:2014ask}.  Additionally, the singlet vev cannot contribute to gauge boson or fermion masses. Hence, we can choose $(v,x)=(v_{\rm EW},0)$, with $v_{\rm EW}=246$~GeV, to be the EW symmetry breaking minimum that recreates the SM mass spectrum. The requirement that this is a minimum of the potential yields.
\begin{eqnarray}
\mu^2=\lambda\,v_{\rm EW}^2\quad{\rm and}\quad b_1=-\frac{v_{\rm EW}^2}{4}a_1.
\end{eqnarray}

After EWSB, there are two scalar mass eigenstates $h_1$ and $h_2$ with masses $m_1$ and $m_2$, respectively, that are admixtures of the scalar bosons $h$ and $s$.
\begin{eqnarray}
\begin{pmatrix} h_1\\h_2\end{pmatrix}=\begin{pmatrix}\cos\theta & \sin\theta\\ -\sin\theta & \cos\theta\end{pmatrix}\begin{pmatrix} h\\s\end{pmatrix}.\label{eq:massmixing}
\end{eqnarray}
Throughout the rest of the paper we will assume $h_1$ is the observed Higgs boson with $m_1=125$~GeV~\cite{ParticleDataGroup:2024cfk} and $h_2$ is a heavier new scalar ($m_2>2\,m_1$).  
Several parameters in the potential can be replaced with the masses $m_1,m_2$ and the scalar mixing angle $\theta$~\cite{Chen:2014ask}:
\begin{eqnarray}
    a_1 &= & \frac{m_1^2 - m_2^2}{v_{\text{EW}}}\sin2\theta, \\
    b_2 &= & m_1^2\sin^2\theta + m_2^2\cos^2\theta - \frac{a_2v_{\text{EW}}^2}{2}, \\
    \lambda &= & \frac{m_1^2\cos^2\theta + m_2^2\sin^2\theta}{2v_{\text{EW}}^2}.\label{eq:lam}
\end{eqnarray}
Including vevs, masses, and mixing angle, the only other free parameters are $a_2$, $b_3$, and $b_4$.

After symmetry breaking, the scalar mass eigenstates have the cubic self-interactions. 
\begin{equation}
    V\supset\frac{\lambda_{111}}{3!}h_1^3 + \frac{\lambda_{112}}{2!}h_2h_1^2 + \frac{\lambda_{122}}{2!}h_2^2h_1 + \frac{\lambda_{222}}{3!}h_2^3.
\end{equation}
The results for the trilinear couplings can be found in Ref.~\cite{Chen:2014ask}.  For di-Higgs production, the relevant trilinear couplings are $\lambda_{111}$ and $\lambda_{112}$.  The former can affect the di-Higgs rate through nonresonant $h_1$ production, $pp\rightarrow h_1^*\rightarrow h_1h_1$, and the latter through resonant $h_2$ production, $pp\rightarrow h_2\rightarrow h_1h_1$.  The partial width for $h_2\rightarrow h_1h_1$ is
\begin{equation}
    \Gamma(h_2 \rightarrow h_1h_1) = \frac{\lambda_{112}^2}{32\pi m_2}\sqrt{1 - \frac{4m_1^2}{m_2^2}}.
\end{equation}

For completeness, the relevant trilinears for $h_1h_1$ production~\cite{Chen:2014ask} are
\begin{eqnarray}
\lambda_{111}&=&3\frac{m_1^2}{v_{\rm EW}}\cos^3\theta+2\,b_3\,\sin^3\theta+3\,a_2\,v_{\rm EW}\cos\theta\sin^2\theta,\label{eq:lam111}\\
    \lambda_{112} &=& \sin\theta\left[-\frac{2m_1^2 + m_2^2}{\vew}\cos^2\theta +2 a_2\vew\left(1 - \frac{3}{2}\sin^2\theta\right) + b_3\sin2\theta\right].
    \label{lambda211}
\end{eqnarray}

\section{Constraints}
\label{constraints}
The real singlet scalar model is subject to various theoretical and experimental constraints.  In this section, we give an overview of the constraints that we impose to determine the maximum di-Higgs rates and variations in the trilinear Higgs coupling.
\subsection{Vacuum stability and perturbative unitarity}
The quartic couplings in the scalar potential are bounded by two conditions: vacuum stability and perturbative unitarity.  For the theory to be viable, the scalar potential needs to be bound so that it does not decay in a runaway direction.  That is, it should be positive at large field values.
\begin{equation}
    \frac{\lambda}{4} h^4 + \frac{a_2}{4}h^2s^2 + \frac{b_4}{4}s^4 > 0.
\end{equation}

Stability along the axes $s = 0$ and $h = 0$ implies that both $\lambda$ and $b_4$ are positive. Additionally,  if $a_2$  is positive, the potential is bound.  However, $a_2$ can also be negative.  To see this, rewrite the quartic terms  as
\begin{equation}
    \frac{\lambda}{4}\left(h^2 + \frac{a_2s^2}{2\lambda}\right)^2 + \frac{1}{4}\left(b_4 - \frac{a_2^2}{4\lambda}\right)s^4.
\end{equation}
The potential is stable for
\begin{equation}
    a_2 \geq -2\sqrt{\lambda b_4}.
\end{equation}

To maintain a perturbative theory, we will also impose perturbative unitarity.  Following from scattering theory applied to QFT, the matrix element of any  two-to-two scalar scattering can be written in terms of the Legendre polynomials as
\begin{equation}
    \mathcal{M} = 16\pi \sum_k \left(2k + 1\right)a_kP_k(\cos\theta),
\end{equation}
where $a_k$ are complex parameters.  
The unitarity bound from the outgoing and incoming waves implies that $a_k$ lies within a circle of radius $1/2$ centered at $i/2$. This leads to the typical perturbative unitarity constraints. 
\begin{eqnarray}
|{\rm Re}(a_k)| \leq 1/2.
\end{eqnarray}
This constraint is usually applied at tree level, where saturation of the limit implies that there is at minimum a 41\% higher order correction to the scattering processes~\cite{Schuessler:2007av}. Since the perturbative unitarity constraint is a limit on higher order corrections, it is a constraint on the perturbativity of the theory.  
 At high energies the matrix elements for two-to-two scalar scattering are
\begin{eqnarray}
    a_0 &=\displaystyle \frac{3b_4}{8\pi} \quad &{\rm for}\quad s\,s\rightarrow s\,s\,, \\
    a_0 &=\displaystyle \frac{a_2}{16 \pi} \quad &{\rm for}\quad s\,h\rightarrow s\,h\,,~{\rm and}\\
    a_0 &= \displaystyle \frac{3 \lambda}{8 \pi} \quad &{\rm for}\quad h\,h\rightarrow h\,h.
\end{eqnarray}
While perturbative unitarity places upper and lower bounds on the quartic couplings, the lower bounds from vacuum stability are more constraining.  The perturbative unitarity and vacuum stability bounds on the quartic couplings are then
\begin{eqnarray}
0\leq b_4 \leq \frac{4\,\pi}{3},\quad 0\leq \lambda \leq \frac{4\,\pi}{3},\quad {\rm and}\quad -2\sqrt{\lambda b_4}\leq a_2 \leq 8\,\pi.~\label{eq:pertunit}
\end{eqnarray}

\subsection{Global minimization}
\begin{figure}[tb]
    \begin{center}
    \subfigure[]{\includegraphics[height=7cm]{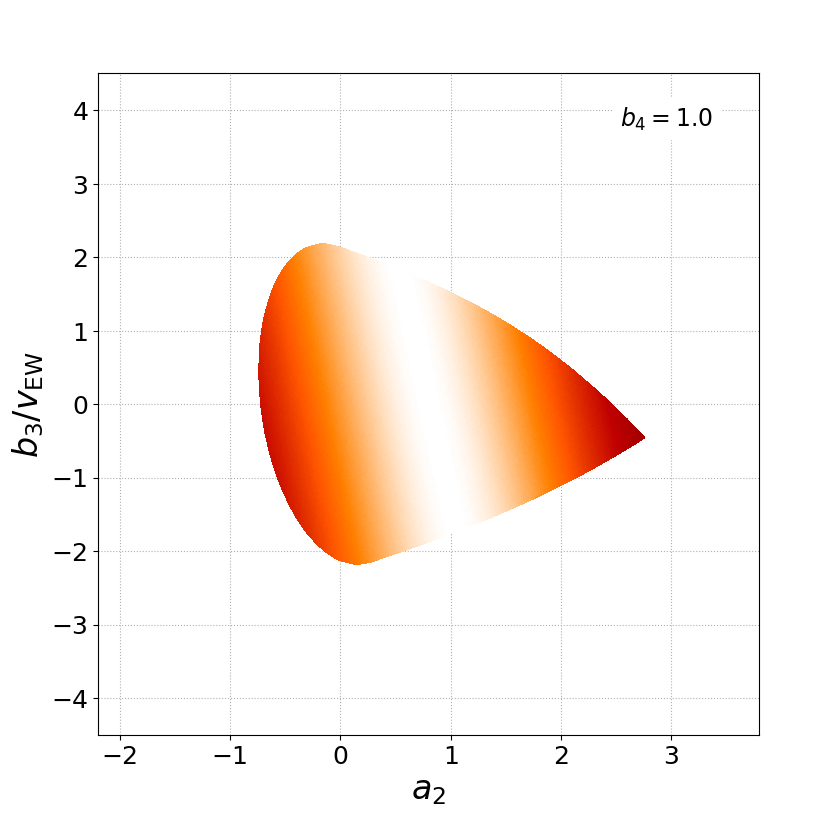}}
    \subfigure[]{\includegraphics[height=7cm]{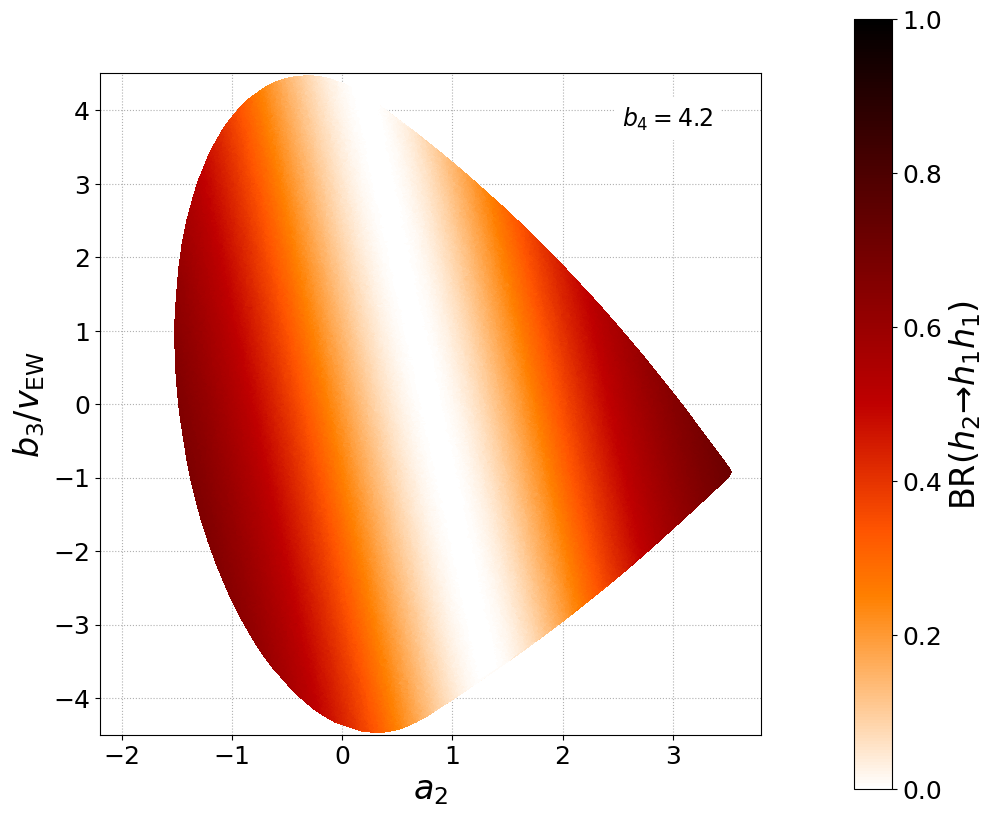}}
    \caption{Heat maps of ${\rm BR}(h_2\rightarrow h_1h_1)$ in the allowed $b_3-a_2$ plane for $m_2 =260$~GeV and $\sin^2\theta = 0.14$. In (a) $b_4 = 1.0$ with a maximum BR of $0.66$, and in (b) $b_4 = 4.2$ with a maximum BR of $0.78$.}
    \label{fig:compareb4}
\end{center}
\end{figure}

The scalar potential in Eq.~(\ref{eq:Potential}) has six possible extrema, with one minimum at $(v,x)=(v_{\rm EW},0)$ by construction. Explicit solutions for $(v,x)$ for all six extrema can be found in Refs.~\cite{Chen:2014ask,Lewis:2017dme}. Since the singlet scalar does not contribute to gauge boson or fermion masses and the other minima do not correspond to a Higgs vev of $v_{\rm EW}$,  the global minimum of the scalar potential must be $(v,x)=(v_{\rm EW},0)$.  This constraint places meaningful bounds on $a_2$ and $b_3$. The global minimum constraint is imposed by checking the value of the scalar potential in Eq.~(\ref{eq:Potential}) at each one of the six extrema.  A parameter point passes if $(v,x)=(v_{\rm EW},0)$ is the lowest lying minimum.  These bounds are relevant for $\lambda_{112}$ and $pp\rightarrow h_2\rightarrow h_1h_1$ rates.  For completeness, in Fig.~\ref{fig:compareb4} we show the allowed region for $a_2$ and $b_3$ under these constraints together with a heat map for ${\rm BR}(h_2\rightarrow h_1h_1)$ for two different choices of $b_4$ and a particular choice of mass and mixing angle. The branching ratio of $h_2\rightarrow h_1h_1$ is maximized on the boundaries of the allowed space, but never reaches one. That is, theoretical constraints on the global minimum and perturbativity of couplings bound how large ${\rm BR}(h_2\rightarrow h_1h_1)$ can be. Some features in Fig.~\ref{fig:compareb4} are fairly generic. The allowed region for $a_2$ and $b_3$ growing as $b_4$ increases is true for all masses and mixing angles. For small enough masses, there will generically be a line with ${\rm BR}(h_2\rightarrow h_1h_1)=0$ somewhere in the allowed region, and there will be two local maxima for ${\rm BR}(h_2\rightarrow h_1h_1)$ along the boundary on either side of this line. For larger masses and nonzero mixing angle, however, the zero branching ratio will no longer be allowed (as one should expect from the Goldstone Boson Equivalence Theorem, which leads to the branching ratio converging to 1/4 as the mass increases), and then the only local maximum for the branching ratio will be the global maximum on the boundary.  Figure~\ref{fig:compareb4} reproduces the results in Refs.~\cite{Chen:2014ask,Lewis:2017dme}, where more details can be found.

\subsection{Experimental}

The singlet model affects precision Higgs rates as well as providing new avenues for discovery via the production of the new scalar.  Higgs precision measurements are typically quoted in terms of signal strengths.
\begin{eqnarray}
\mu^i_f=\frac{\sigma_i(pp\rightarrow h_1)}{\sigma_{i,\rm SM}(pp\rightarrow h_1)}\frac{{\rm BR}(h_1\rightarrow f)}{{\rm BR}_{\rm SM}(h_1\rightarrow f)},
\end{eqnarray}
where the initial state is $i$, the final state $f$, and the subscript ${\rm SM}$ indicates SM values.  In the renormalizable real singlet model with $m_2>m_1$, from the mixing in Eq.~(\ref{eq:massmixing}), all production and decay rates for $h_1$ are universally suppressed by $\cos^2\theta$ compared to the SM rates:
\begin{eqnarray}
\sigma_i(pp\rightarrow h_1)=\cos^2\theta\,\sigma_{i,\rm SM}(pp\rightarrow h_1)\quad{\rm and}\quad \Gamma(h_1\rightarrow f)=\cos^2\theta\,\Gamma_{\rm SM}(h_1\rightarrow f).
\end{eqnarray}
The branching ratios of $h_1$ are unchanged from SM predictions since they are ratios of rates. Hence, there is a universal signal strength for all production and decay channels.
\begin{eqnarray}
\mu^i_f = \cos^2\theta.\label{eq:SigStrength}
\end{eqnarray}
That is, measurements of signal strengths will constrain $\theta$ independently of the mass of $h_2$.

Meanwhile, the couplings of $h_2$ to SM gauge bosons and fermions are suppressed by $\sin\theta$ relative to the SM prediction for Higgs couplings.  Hence, the production rates from and decay rates into SM gauge bosons and fermions are
\begin{equation}
\sigma_i(pp\rightarrow h_2)=\sin^2\theta\,\sigma_{i,\rm SM}(pp\rightarrow h_2)\quad {\rm and}\quad
\Gamma(h_2 \to f_{\text{SM}}) = \sin^2\theta\,\Gamma_{\text{SM}}(h_2 \to f_{\text{SM}}),\label{eq:h2Rates}
\end{equation}
where $f_{\rm SM}$ are final state SM gauge bosons or fermions, and the subscript ${\rm SM}$ refers to SM Higgs predictions at the mass of $h_2$.  Above the di-Higgs threshold the new decay $h_2\rightarrow h_1h_1$ opens up and the total width for $h_2$ is then given by
\begin{equation}
    \Gamma(h_2) = \Gamma(h_2 \to h_1h_1) + \Gamma(h_2 \to f_{\text{SM}}).
\end{equation}
The total rate is then approximated using the narrow width approximation.
\begin{equation}
    \sigma_i(pp\to h_2 \to f) \approx \sigma_i(pp \to h_2)\text{BR}(h_2 \to f).
\end{equation}
To ensure that the narrow width approximation is valid, we demand that the total $h_2$ width is less than 10\% of its mass.
\begin{equation}
    \Gamma(h_2) \leq 0.1 m_2.\label{eq:NWA}
\end{equation}
Such a constraint also helps keep the theory perturbative.

\begin{figure}[t]
    \centering
    \subfigure[]{\includegraphics[width=0.45\textwidth,clip]{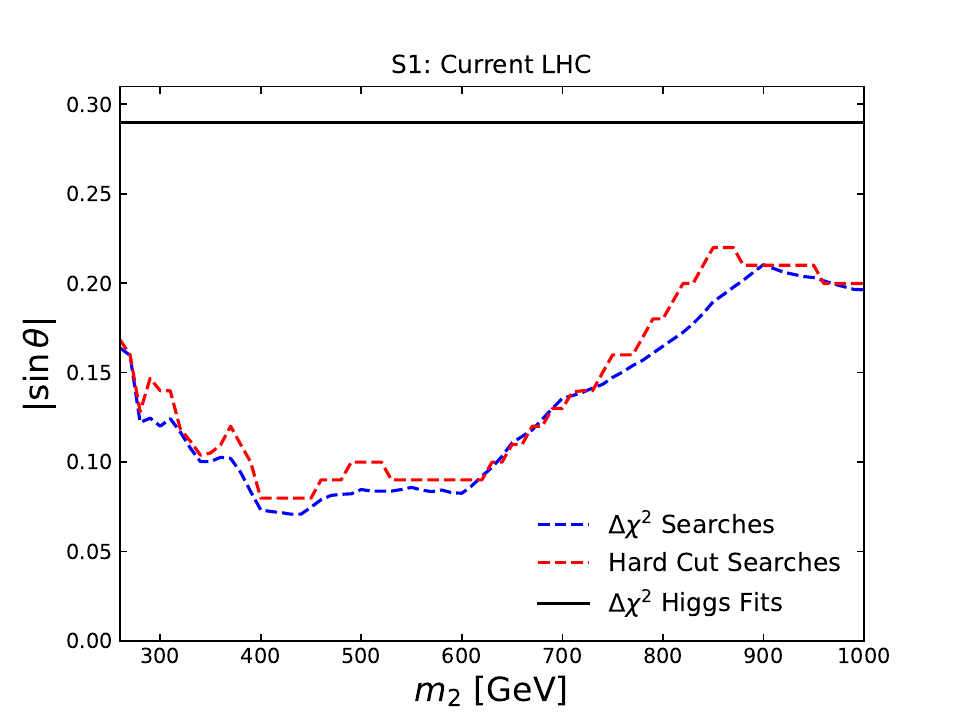}}
    \subfigure[]{\includegraphics[width=0.45\textwidth,clip]{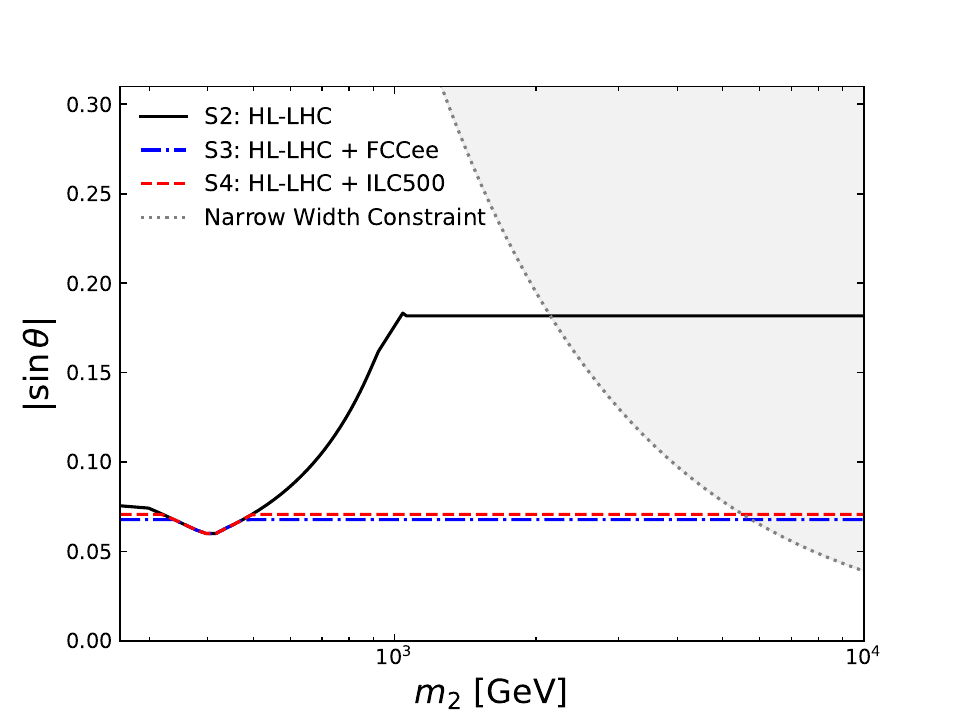}}
    \caption{(a) Mixing angle constraints for scenario S1 from (black solid) Higgs precision measurement, (red dashed) hard cut bounds from scalar searches, and (blue dashed) a $\Delta\chi^2$ combination of scalar searches and Higgs precision measurements.  (b) Mixing angle constraints for scenarios (black solid) S2, (blue dot dashed) S3, and (red dashed) S4.  The shaded region is excluded by the narrow width requirement.}
    \label{fig:Max_sin}
\end{figure}

In Fig.~\ref{fig:Max_sin}, we show the upper bounds on the scalar mixing angle $|\sin\theta\,|$ for scenarios S1-S4 as described in the Introduction.  For current LHC constraints from Higgs precision and direct scalar searches, we use the results of Ref.~\cite{Lane:2024vur}.  The dominant search channels are $h_2\rightarrow W^\pm W^\mp$, $h_2\rightarrow ZZ$, and $h_2\rightarrow h_1h_1$.  These are the largest branching ratios in the real singlet model, since according to Eq.~(\ref{eq:h2Rates}) $h_2$ decays similarly to a SM like Higgs at mass $m_2$.  As shown in Ref.~\cite{Lane:2024vur}, for $220 <m_2 < 250$~GeV, $m_2 \approx 280$~GeV, $320 < m_2 < 350$~GeV the resonance searches for $WW$ and $ZZ$ are the most constraining. Outside these regions, the resonant di-Higgs searches are most constraining.  These constraints have two scenarios for combining precision Higgs measurements with scalar searches: a ``hard cut'' and ``$\Delta\chi^2$'' method~\cite{Bechtle:2008jh,Adhikari:2020vqo,Bahl:2022igd,Lane:2024vur}.  We consider both results. Reference~\cite{Feuerstake:2024uxs} considers the limits derived from HiggsTools~\cite{Bahl:2022igd}; the methods we use, which include newer experimental results, are more constraining. For the HL-LHC, FCC-ee, and ILC500 we use the projections from the 2019 European Strategy Report~\cite{EuropeanStrategyforParticlePhysicsPreparatoryGroup:2019qin} and Snowmass 2022~\cite{Dawson:2022zbb}. As discussed previously around Eq.~(\ref{eq:SigStrength}), Higgs precision measurements constrain $\sin\theta$ independent of the $h_2$ mass.  Hence, flat regions in Fig.~\ref{fig:Max_sin} arise from fitting Higgs data.  At masses around or below 1 TeV, scalar searches can dominate leading to a mass dependent $\sin\theta$ bound.  At the LHC scalar searches are always more constraining, while for the other scenarios Higgs precision can be more constraining.  

The constraint from the narrow width is also shown as a shaded region in Fig.~\ref{fig:Max_sin}. At high masses, the decays $h_2\rightarrow W^\pm W^\mp$, $h_2\rightarrow ZZ$, and $h_2\rightarrow h_1h_1$ dominate. Using the Goldstone Boson Equivalence Theorem as well as taking the high-mass limit of the SM-like widths of $h_2\rightarrow W^\pm W^\mp$ and $h_2\rightarrow ZZ$ with mixing angle suppression, the total width of $h_2$ is then 
\begin{eqnarray}
\Gamma(h_2)\approx \frac{4}{3}\sin^2\theta\,\Gamma_{SM}(h_2)\approx \frac{m_2^3}{8\,\pi v_{\rm EW}^2}\sin^2\theta.\label{eq:GBET}
\end{eqnarray}
A more detailed derivation of Eq.~(\ref{eq:GBET}) can be found in Appendix.~\ref{app:GBET}.
Equation~(\ref{eq:NWA}) leads to a bound on $\sin\theta$ of
\begin{equation}
    \sin^2\theta \lesssim \frac{0.8\pi\vew^2}{m_2^2}.
    \label{eq:sin_narrow-width}
\end{equation}
This is a monotonically decreasing upper bound on $\sin\theta$ for large $m_2$. The narrow width constraint from Eq.~(\ref{eq:sin_narrow-width}) is shown as the shaded grey region in Fig.~\ref{fig:Max_sin}(b).  The narrow width constraint is dominant over experimental limits for the HL-LHC when $m_2\gtrsim2.1$ TeV and for HL-LHC+FCCee and HL-LHC+ILC500 when $m_2\gtrsim5.7$~TeV.

There are also constraints from EW precision observables on the mixing between a scalar singlet and the SM Higgs doublet~\cite{Profumo:2007wc,Dawson:2009yx,Chen:2014ask,Lopez-Val:2014jva,Falkowski:2015iwa,Dawson:2020oco,Papaefstathiou:2022oyi}. When considered by itself, constraints from the $W$ mass~\cite{Lopez-Val:2014jva,Papaefstathiou:2022oyi} ($|\sin\theta|\lesssim 0.15$ for $m_2\sim 1$~TeV) are stronger than LHC signal strengths or direct searches. However, when the model is fit to all EW precision observables, these constraints are weakened~\cite{Lopez-Val:2014jva,Falkowski:2015iwa,Dawson:2020oco}.  A recent fit to EW precision observables~\cite{Dawson:2020oco} found $|\sin\theta|\lesssim 0.2$ at $m_2\sim1$~TeV and weaker for lower masses.  These are similar to or weaker than the LHC bounds in Fig.~\ref{fig:Max_sin}(a).  Hence, we use the bounds in Fig.~\ref{fig:Max_sin} in this work.

\section{Results}
\label{sec:Results}
We now provide the main results of this paper: finding the maximum allowed resonant di-Higgs rate and Higgs trilinear couplings in the real singlet extension.  As discussed in the Introduction, we consider four scenarios for experimental constraints from different colliders
\begin{enumerate}[label=(\roman*)]
\item S1: Current LHC~\cite{Lane:2024vur}.
\item S2: The HL-LHC~\cite{EuropeanStrategyforParticlePhysicsPreparatoryGroup:2019qin,Dawson:2022zbb}.
\item S3: The HL-LHC+FCCee~\cite{EuropeanStrategyforParticlePhysicsPreparatoryGroup:2019qin,Dawson:2022zbb}. 
\item S4: The HL-LHC+ILC500~\cite{EuropeanStrategyforParticlePhysicsPreparatoryGroup:2019qin,Dawson:2022zbb}.  
\end{enumerate}
All constraints in Sec.~\ref{constraints}, including all theoretical constraints, are applied to a scan over $\sin\theta$, $a_2$, $b_3$, $b_4$, and $m_2$.  

\begin{figure}[t]
    \centering
        \subfigure[]{\includegraphics[width=0.45\textwidth,clip]{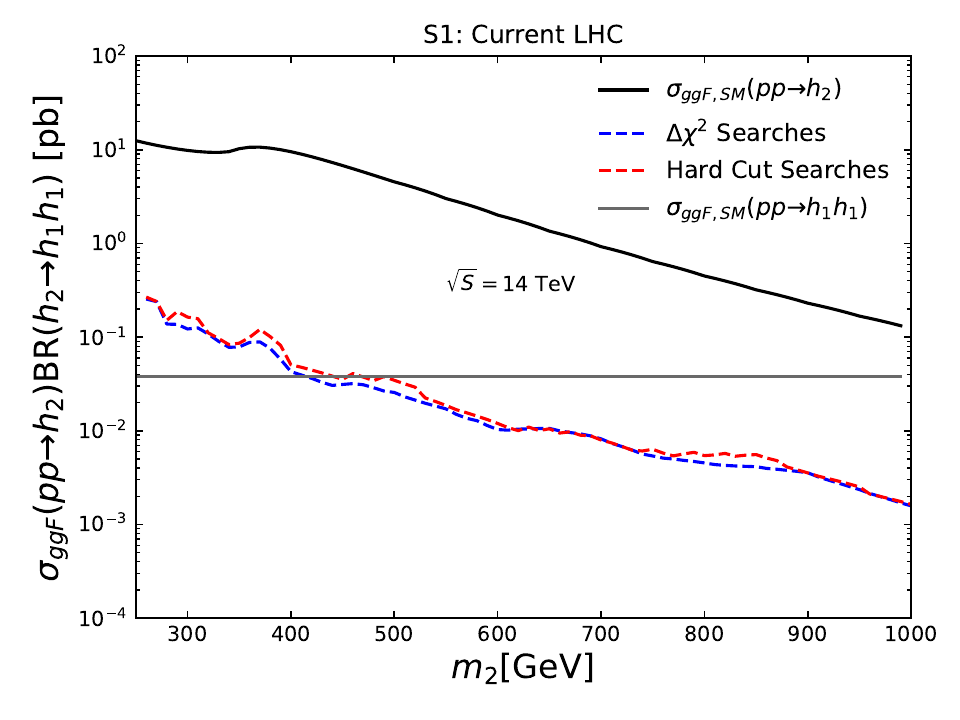}}
    \subfigure[]{\includegraphics[width=0.45\textwidth,clip]{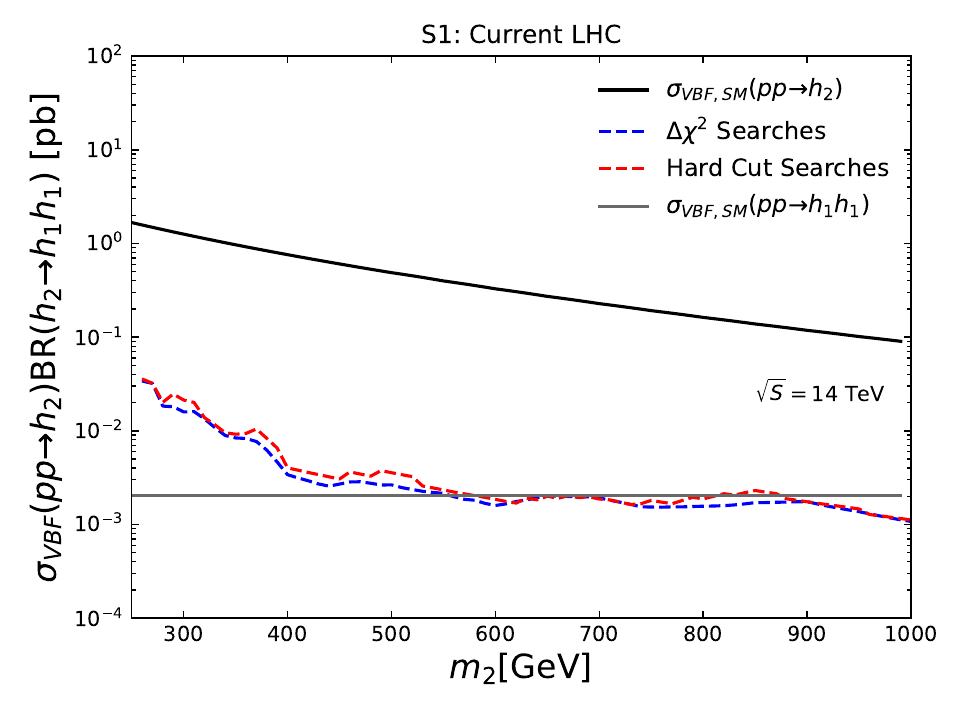}}
    \subfigure[]{\includegraphics[width=0.45\textwidth,clip]{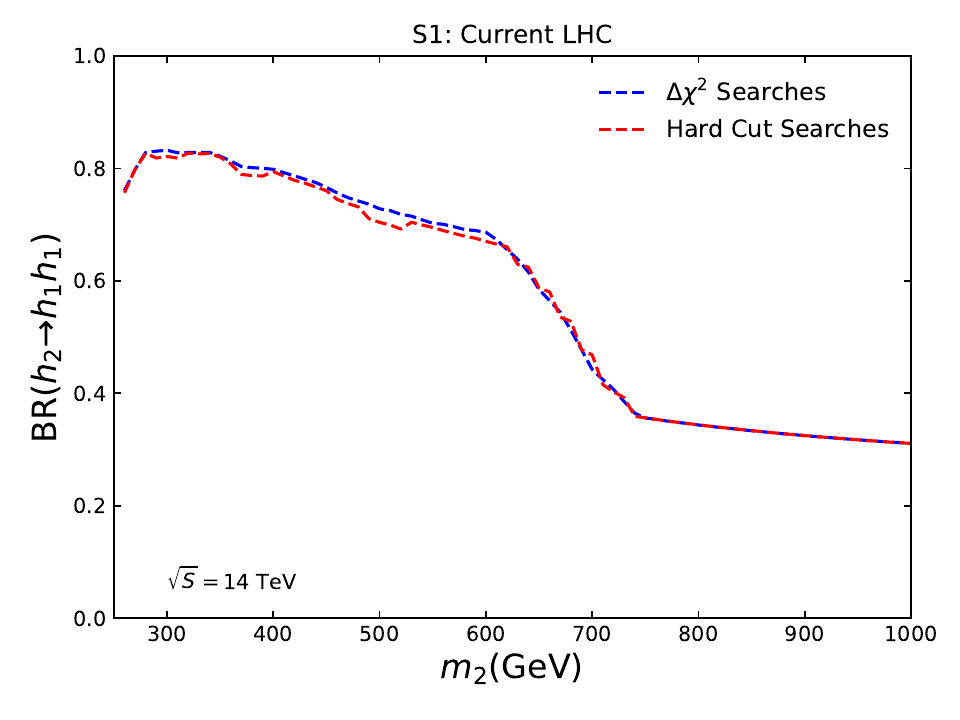}}
    \caption{Maximum allowed cross section $\sigma(pp \to h_2 \to h_1h_1)$ in scenario S1 at lab frame energy $\sqrt{S}=14$~TeV for (a) gluon fusion and (b) VBF, with (c) the corresponding branching ratio ${\rm BR}(h_2\rightarrow h_1h_1)$.  Two possibilities for $\sin\theta$ limits from Higgs precision and scalar searches are shown: blue dashed lines for the $\Delta\chi^2$ combination and red dashed for the hard cut limits.  In (a,b) the light grey line is for the SM di-Higgs rate and the black solid the SM production rate at mass $m_2$ for (a) gluon fusion and (b) VBF.}
    \label{fig:cuts}
\end{figure}

In Fig.~\ref{fig:cuts} we show the maximum allowed resonant di-Higgs rates in scenario S1 for the (a) gluon fusion and (b) vector boson fusion (VBF) production modes of $h_2$ together with (c) the associated branching ratio ${\rm BR}(h_2\rightarrow h_1h_1)$. For all SM-like production rates $\sigma_{i,{\rm SM}}(pp\rightarrow h_2)$, we use the NNLO+NNLL QCD results for gluon fusion and NNLO QCD results for VBF from the LHC Higgs Cross Section Working Group~\cite{LHCHiggsCrossSectionWorkingGroup:2016ypw}.  The current LHC precision Higgs measurements are combined with scalar searches using a traditional (red dashed) ``hard cut'' method~\cite{Bechtle:2008jh,Adhikari:2020vqo,Bahl:2022igd,Lane:2024vur} as well as a (blue dashed) ``$\Delta \chi^2$''~\cite{Adhikari:2020vqo, Lane:2024vur} combination.  As can be seen, both methods have very similar results.  For comparison we have included the predicted SM rates for (black solid) a SM Higgs produced at mass $m_2$, and the (grey solid) SM di-Higgs rates.  As can be seen, at the di-Higgs threshold the gluon fusion rate can still be $\sim 5-10$ times larger than SM di-Higgs, and the VBF rate can be $\sim 15-20$ times larger.  At larger masses near 1 TeV, the maximum gluon fusion rate is over an order of magnitude less than the SM di-Higgs rate while the maximum VBF rate is only a factor of two less than the SM rate.  Although the $\sin\theta$ limit does not change much between $m_2\sim250-1000$~GeV, the SM cross section at $m_2$ and the ${\rm BR}(h_2\rightarrow h_1h_1)$ do decrease with mass.  The branching ratio of $h_2\rightarrow h_1h_1$ corresponding to the maximum rate is near 80\% near the di-Higgs threshold.  However, as  $m_2$ increases, ${\rm BR}(h_2\rightarrow h_1h_1)$ decreases approaching 1/4 as dictated by the Goldstone Boson Equivalence Theorem.  

\begin{figure}
    \centering
    \subfigure[]{\includegraphics[width=0.45\textwidth,clip]{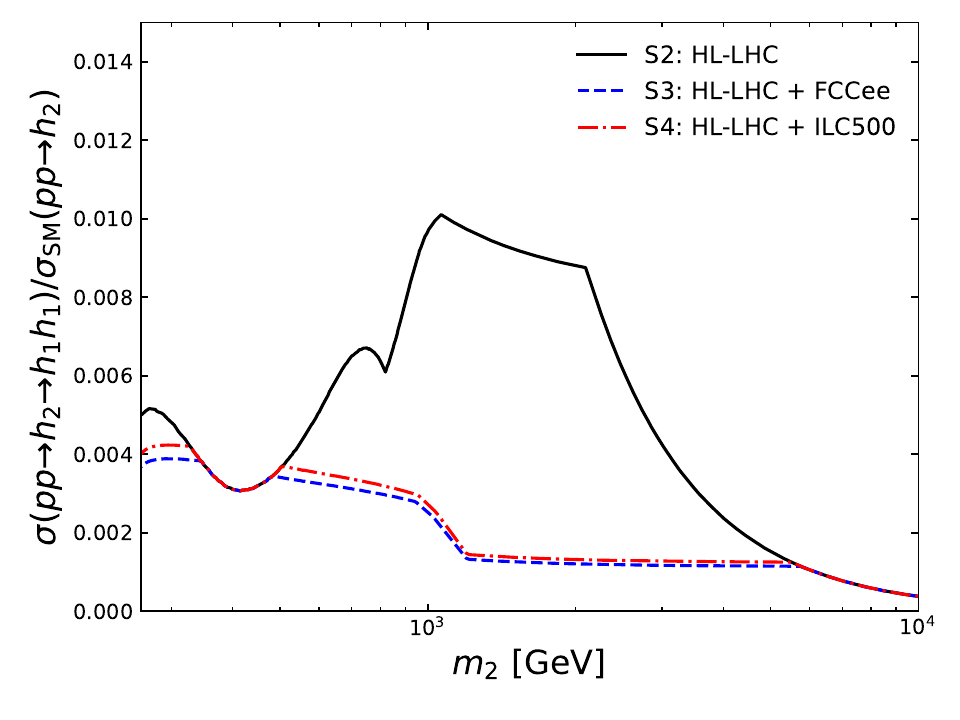}}
    \subfigure[]{\includegraphics[width=0.45\textwidth,clip]{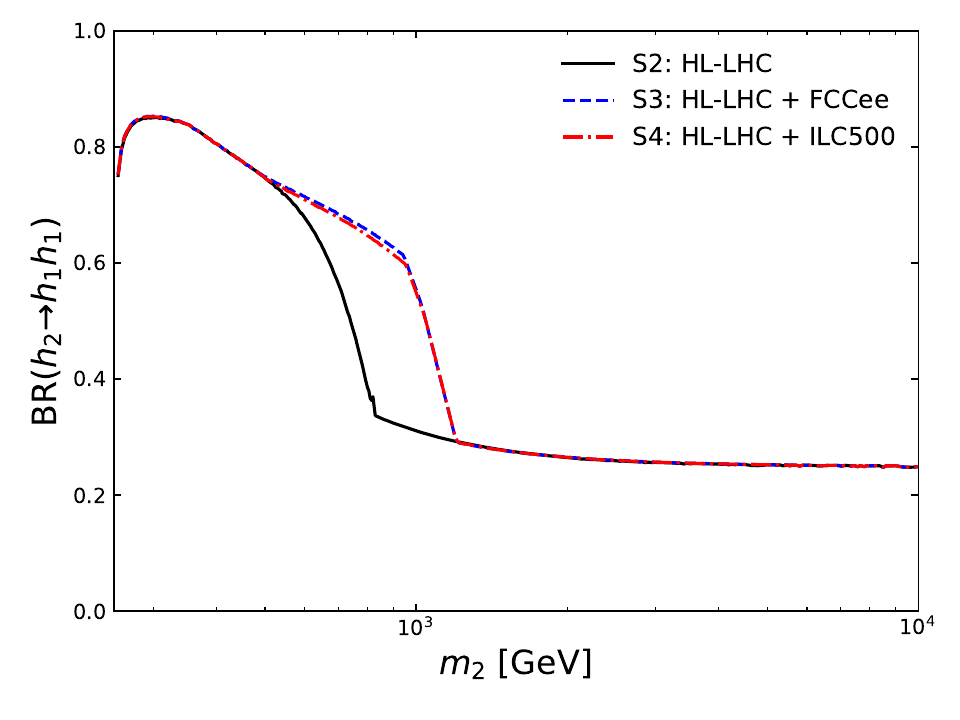}}
    \subfigure[]{\includegraphics[width=0.45\textwidth,clip]{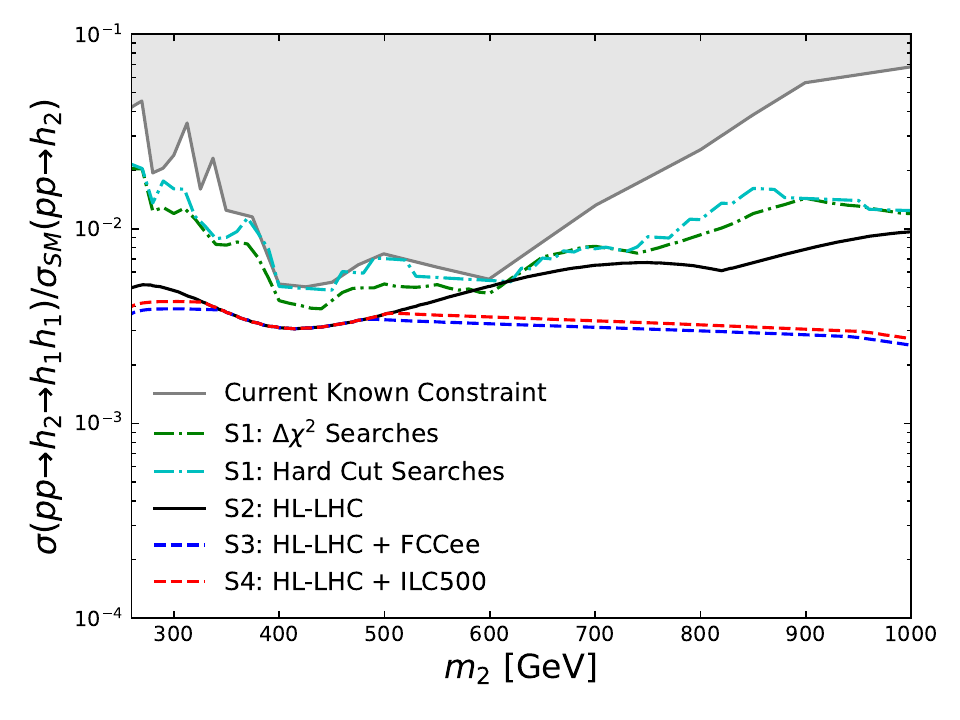}}
    \subfigure[]{\includegraphics[width=0.45\textwidth,clip]{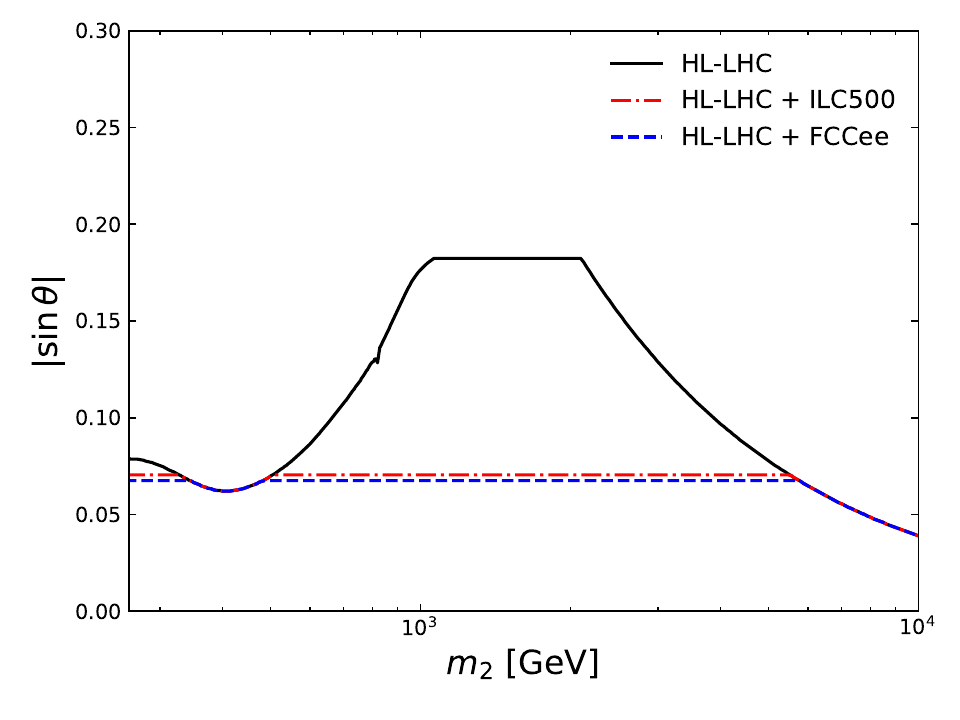}}
    
    \caption{(a) Normalized maximum double Higgs resonant production rates and (b) the corresponding branching ratio ${\rm BR}(h_2\rightarrow h_1h_1)$ for scenarios (black solid) S2, (blue dashed) S3, and (red dot-dashed) S4. In (c) the maximum rates are compared to (grey solid) current di-Higgs upper bounds. (d) The values of $|\sin\theta|$ corresponding to maximum rates.}
    \label{fig:BR}
\end{figure}

Scenarios S2-S4 are projections for future colliders such as a 100 TeV pp collider~\cite{Tang:2015qga,CEPC-SPPCStudyGroup:2015csa,FCC:2018vvp}. Since it is unknown which future high energy colliders may be built, we present the results normalized to the SM prediction for the production of the SM Higgs boson at mass $m_2$.
\begin{eqnarray}
\frac{\sigma(pp\to h_2 \to h_1h_1)}{\sigma_{\text{SM}}(pp \to h_2)} \approx \sin^2\theta\,\text{BR}(h_2 \to h_1h_1).
\end{eqnarray}
These rates can then be translated to any collider in which $h_2$ is singly produced via couplings to SM vector boson or fermions via renormalizable interactions.

In Fig.~\ref{fig:BR} the (a) normalized maximum resonant di-Higgs rate and (b) the corresponding branching ratios for scenarios S2-S4 are shown.  For masses below $1$~TeV, these projected maximum rates are compared to the current constraints on di-Higgs resonance rates in Fig.~\ref{fig:BR}(c).  As can be seen, the projected  maximum rates are still below current constraints, although, the current limit on resonant di-Higgs production is very close to the maximum rate for the HL-LHC projection at $m_2\sim 600$~GeV. This occurs since future projections assume that the measured rates agree with the SM predictions while current searches have statistical fluctuations. Indeed, at $m_2\sim 600$~GeV, there was a downward fluctuation in the di-Higgs limits in Ref.~\cite{ATLAS-CONF-2021-052}. For completeness, we also give the values of $|\sin\theta|$ corresponding to the maximum rates in Fig.~\ref{fig:BR}(d).

The projected maximum rates and corresponding branching ratios in these scenarios have some interesting behaviors.  First, we discuss the HL-LHC (S1) results.  The maximum rate decreases from $m_2\sim 275$~GeV to 425 GeV then starts increasing with the increasing $|\sin\theta|$ limit in Fig.~\ref{fig:Max_sin}(b).  Around $m_2\sim750-820$~GeV, the projected maximum rate decreases and then increases again.  For $h_2$ masses near but below $500$ GeV, the maximum ${\rm BR}(h_2\rightarrow h_1h_1)$ decreases more quickly than the bound on $|\sin\theta|$ increases, causing the maximum rate to decrease. Between $500$ and $750$ GeV, the bounds on $\sin\theta$ increase more quickly than the decrease in branching ratio. At $750$~GeV, the branching ratio of $h_2\rightarrow h_1h_1$ drops precipitously causing a drop in the allowed maximum rate.  Above $m_2\sim 820$ GeV, the change in the $h_2\rightarrow h_1h_1$ branching ratio flattens.  The maximum production rate then starts increasing quickly with $|\sin\theta|$ until $m_2\sim 1.1$~TeV when the $|\sin\theta|$ bound is from precision Higgs constraints and is flat. Between $m_2\sim1.1-2.1$ TeV, the maximum rate decreases with the slow decrease in ${\rm BR}(h_2\rightarrow h_1h_1)$.  For masses $m_2\gtrsim 2.1$~TeV, the bound on $|\sin\theta|$ decreases quickly due to the narrow width constraint as shown in Eq.~(\ref{eq:sin_narrow-width}). Hence, the maximum rate also decreases quickly.

The behavior of the maximum rates and corresponding branching ratios for the future electron-positron collider scenarios HL-LHC+FCCee (S3) and HL-LHC+ILC500 (S4) are similar to HL-LHC (S1), but with different mass scales.  For masses up to $m_2\sim 1.2$~TeV, the total rate decreases due to a flat upper bound on $|\sin\theta|$ and decreasing branching ratio.  This behavior is only interrupted when the HL-LHC direct scalar search bounds are more constraining than the precision Higgs constraints for $m_2\sim350-500$~GeV.   Around $m_2\sim 1$~TeV the $h_2\rightarrow h_1h_1$ branching ratio corresponding to the maximum rate drops quickly and then flattens out for $m_2\gtrsim 1.2$~TeV.  Hence, between $\sim 1.2$~TeV and $5.7$~TeV the maximum rate is flat corresponding to a flat $|\sin\theta|$ bound and mostly flat $h_2\rightarrow h_1h_1$ branching ratio.  Above $5.7$ TeV, the narrow width constraint of Eq.~(\ref{eq:sin_narrow-width}) sets in, and the maximum rate decreases with the $|\sin\theta|$ limit.

The maximum rate in all cases is continuous, although there are kinks.  However, in scenarios S2-S4, the corresponding branching ratio for $h_2\rightarrow h_1h_1$ and $|\sin\theta|$ have some interesting behavior.  There are clearly two different regions for the branching ratios: a slight increase above threshold with a decrease soon after, and then at $m_2\sim800$~GeV for the HL-LHC (S2) and $m_2\sim 1.1-1.2$~TeV for HL-LHC+ILC500 (S4) and HL-LHC+FCCee (S3), the branching ratios drops precipitously and then flattens out.  This is particularly striking at the HL-LHC, where the branching ratio for $h_2\rightarrow h_1h_1$ and $|\sin\theta|$ (see Figs.~\ref{fig:BR}(b,d)) have a discontinuity at $m_2\approx 800$~GeV. Since some of the details are rather involved, we refer the interested reader to Appendix ~\ref{app:kinks} for detailed explanations for the behaviors. Ultimately, the features can be understood from the interplay between the different parameter constraints and the rate maximization procedure.

\subsection{Variation in the Higgs trilinear coupling}

\begin{figure}[t]
    \centering
      \subfigure[]{\includegraphics[width=0.45\textwidth,clip]{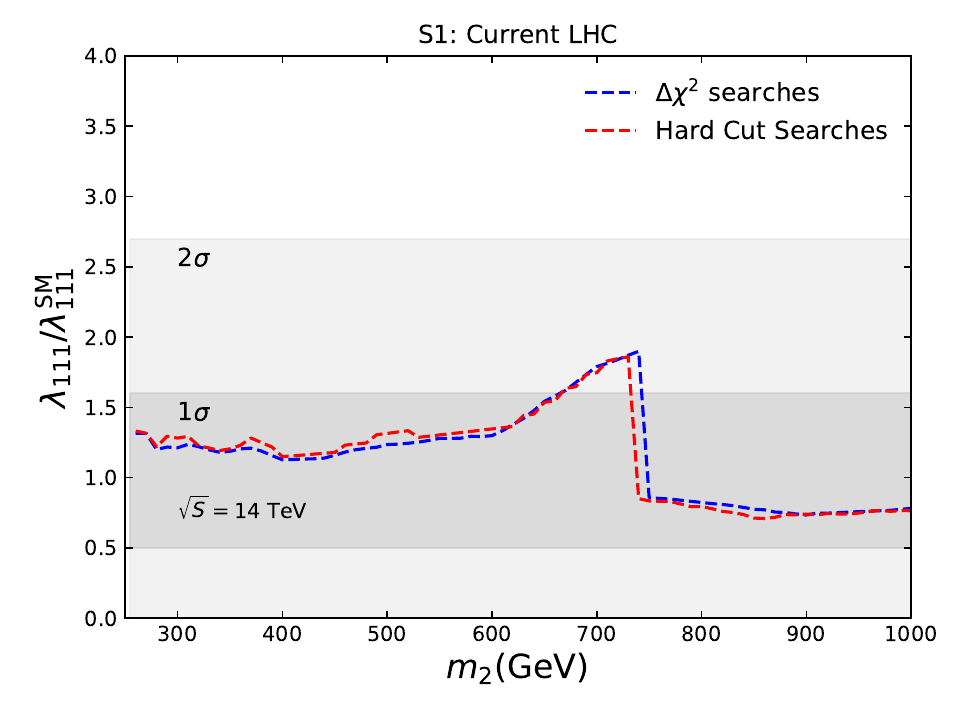}}
   \subfigure[]{\includegraphics[width=0.45\textwidth,clip]{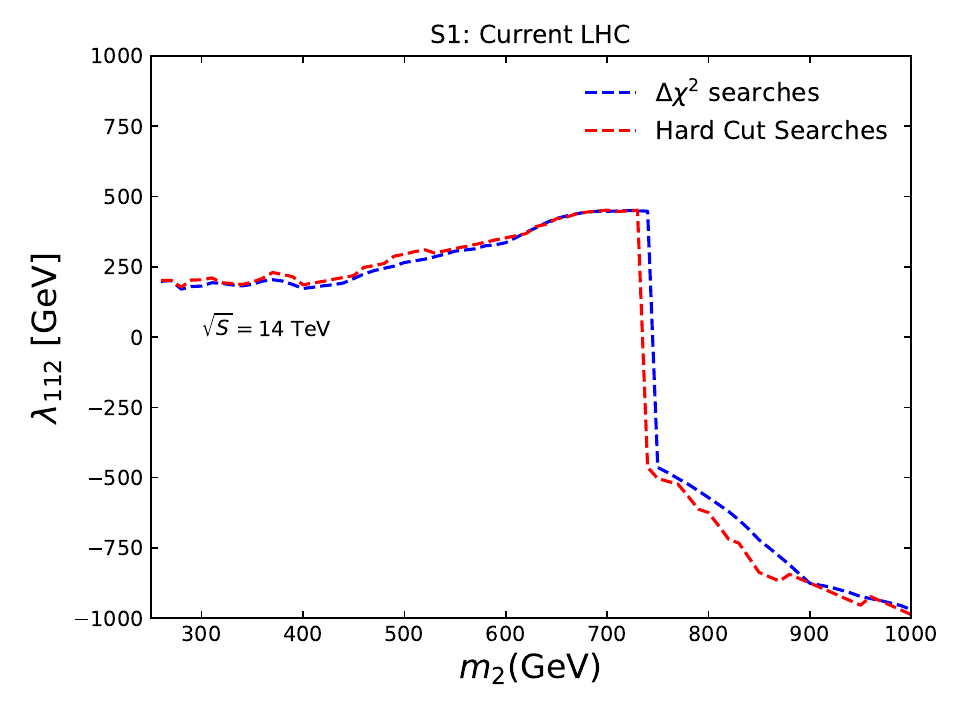}}
      \subfigure[]{\includegraphics[width=0.45\textwidth,clip]{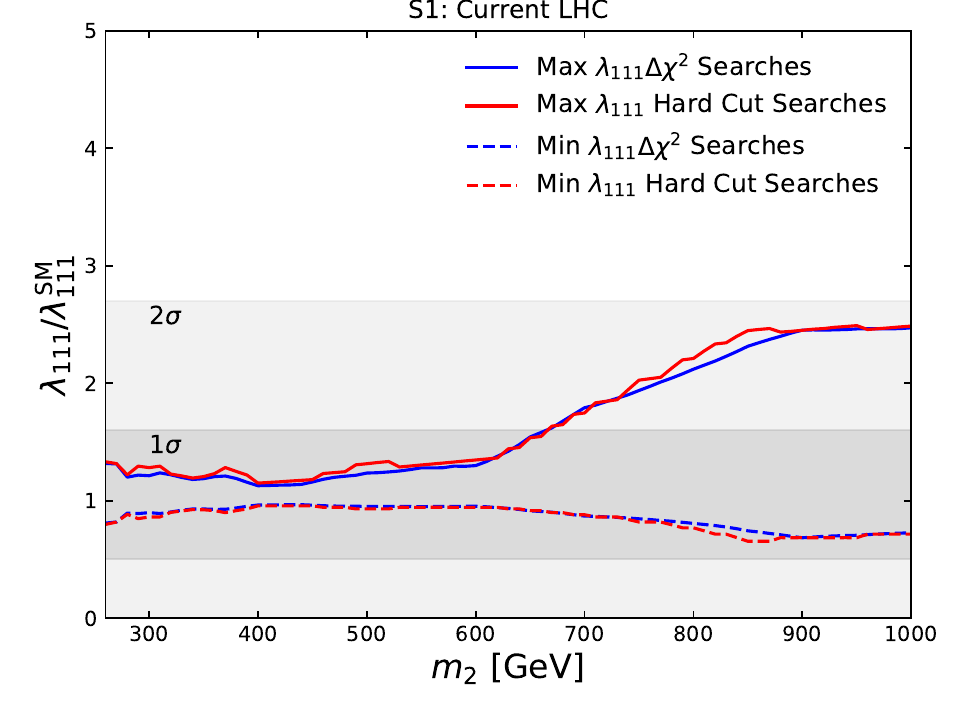}}
   \subfigure[]{\includegraphics[width=0.45\textwidth,clip]{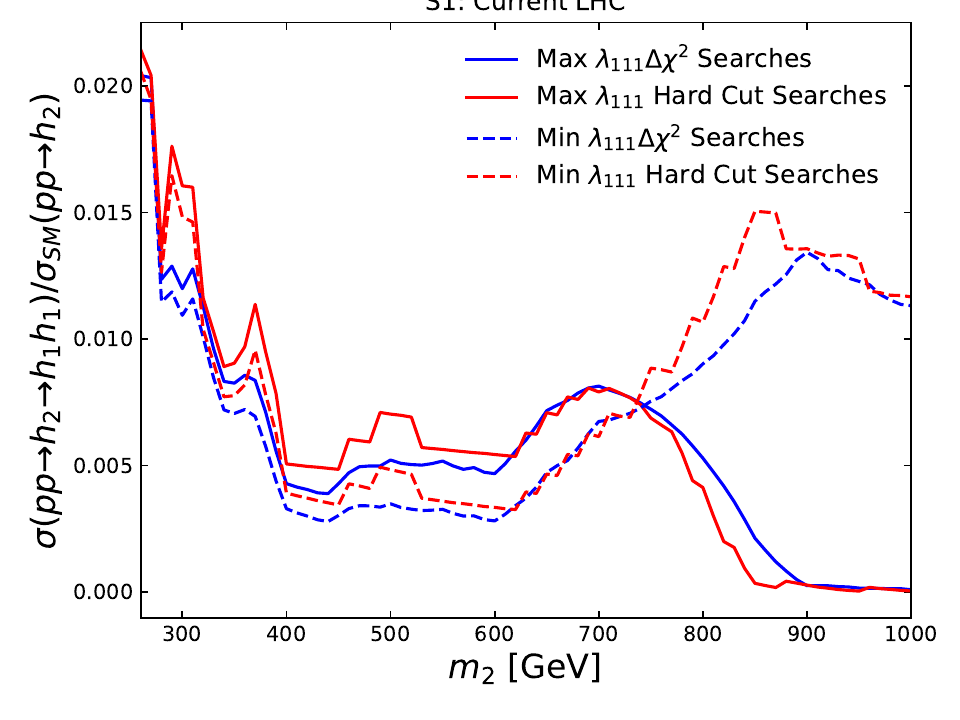}}
    \caption{Values of (a) $\lambda_{111}$ normalized to the SM prediction $\lambda_{111}^{\rm SM}$ and (b) $\lambda_{112}$ corresponding to the maximum di-Higgs rate.  The (solid) maximum and (dashed) minimum allowed $\lambda_{111}$ are shown in (c), and the corresponding $pp\rightarrow h_2\rightarrow h_1h_1$ rates in (d).   All results are presented for current LHC constraints using the (blue) $\Delta\chi^2$ and (red) hard cut combinations of Higgs precision scalar searches. In (a,c) the 1$\sigma$ and 2$\sigma$ projected bounds on $\lambda_{111}$ at the HL-LHC~\cite{ATL-PHYS-PUB-2022-005} are shown in dark and light grey shaded regions, respectively.}
    \label{fig:lambdas_MaxSM}
\end{figure}

We now discuss the variation of the Higgs trilinear coupling $\lambda_{111}$ in this model fully accounting for theoretical constraints and in scenarios S1-S4. In Fig.~\ref{fig:lambdas_MaxSM} we show the values of (a) $\lambda_{111}$ and (b) $\lambda_{112}$ when the $pp\rightarrow h_2\rightarrow h_1h_1$ rate is maximized using current LHC limits.  While for most of the benchmark points, $\lambda_{111}$ is within $\sim 50\%$ of the SM prediction, it can be enhanced by upwards of a factor of two for $m_2\sim700$~GeV.  In Fig.~\ref{fig:lambdas_MaxSM}(c) we show maximum and minimum allowed values of $\lambda_{111}$ considering all theoretical and current LHC constraints, and in (d) the corresponding $pp\rightarrow h_2\rightarrow h_1h_1$ rate.  With current constraints, the $\lambda_{111}$ can still be a factor 2.5 larger than the SM value. For comparison, the 1$\sigma$ and 2$\sigma$ projected bounds on $\lambda_{111}$ at the HL-LHC are also shown~\cite{ATL-PHYS-PUB-2022-005}. The maximum allowed values of $\lambda_{111}$ are less than the projected 2$\sigma$ bounds.  A similar study was done in the $Z_2$ symmetric real singlet model~\cite{DiLuzio:2017tfn}. That study found larger variations in $\lambda_{111}$, but had weaker constraints on $\sin\theta$ due to being conducted several years ago.  The current, stronger limits on the scalar mixing angle limit how large the Higgs trilinear coupling can be.

\begin{figure}[t]
    \centering
      \subfigure[]{\includegraphics[width=0.45\textwidth,clip]{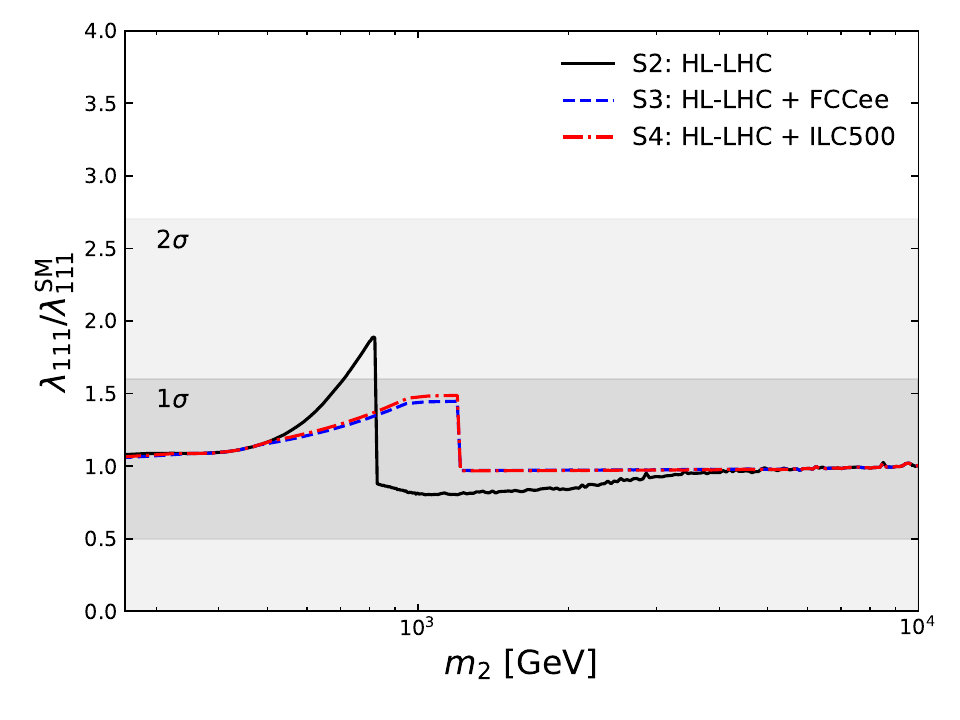}}
   \subfigure[]{\includegraphics[width=0.45\textwidth,clip]{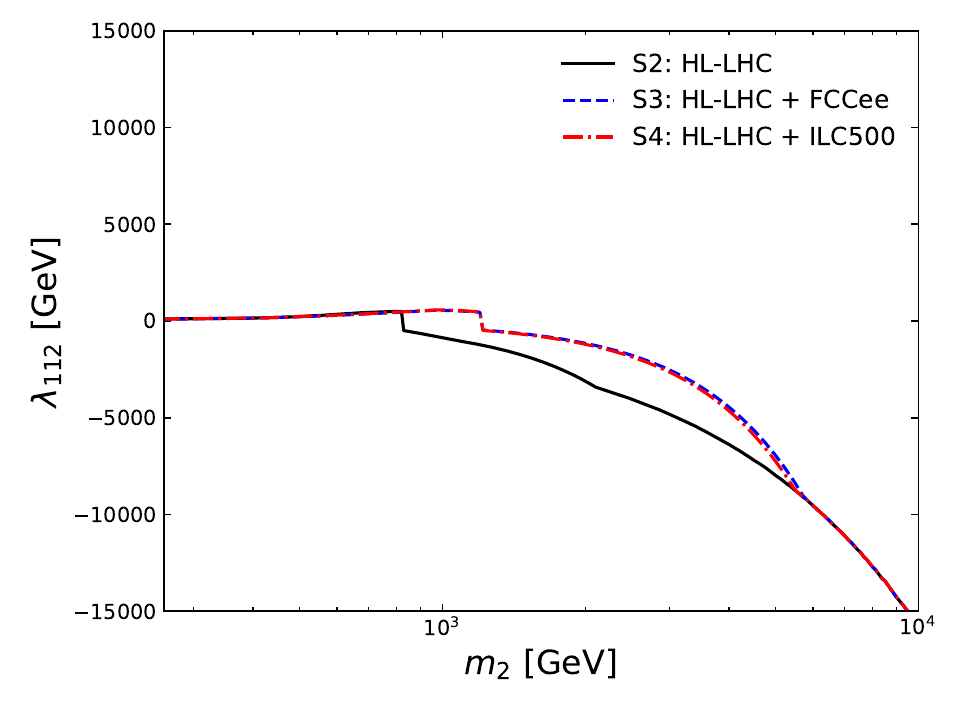}}
      \subfigure[]{\includegraphics[width=0.45\textwidth,clip]{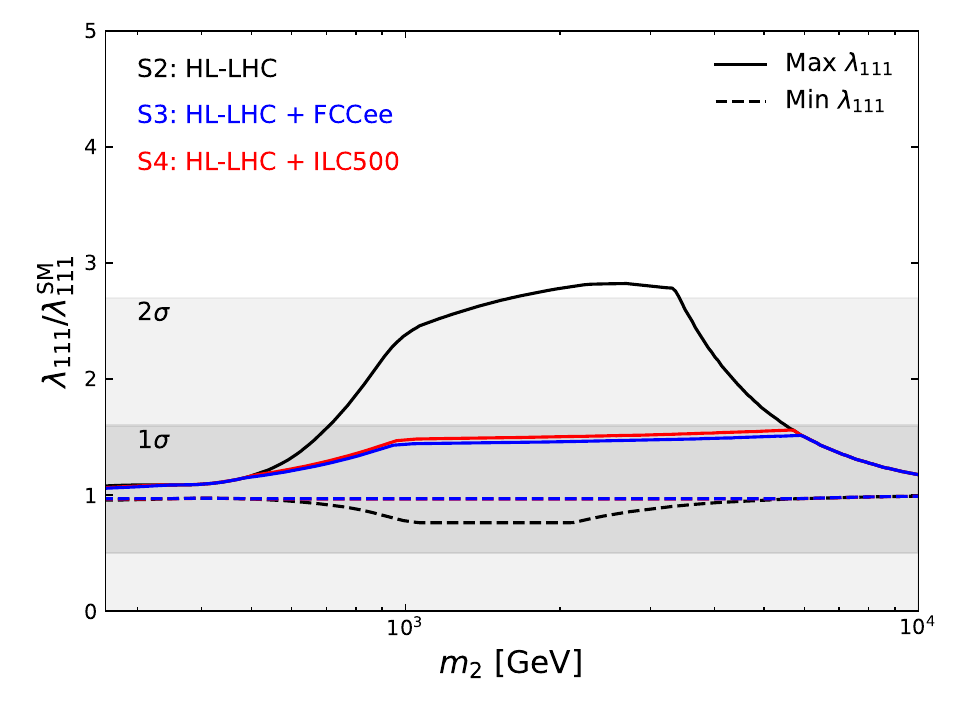}}
   \subfigure[]{\includegraphics[width=0.45\textwidth,clip]{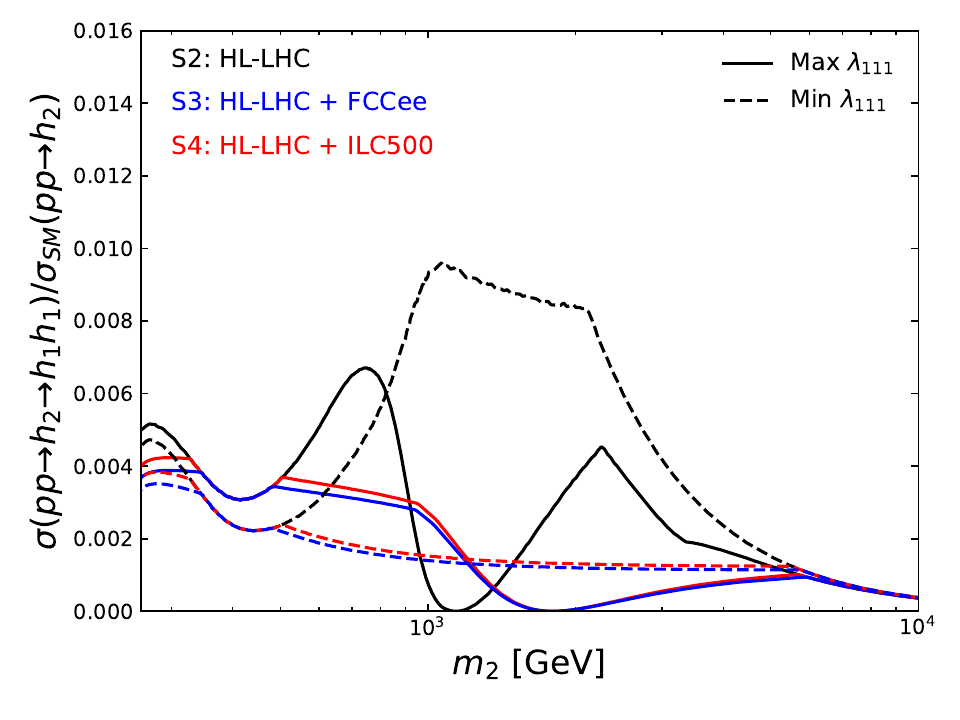}}
    \caption{Values of (a) $\lambda_{111}$ normalized to the SM prediction $\lambda_{111}^{\rm SM}$ and (b) $\lambda_{112}$ corresponding to the maximum di-Higgs rate.  The (solid) maximum and (dashed) minimum allowed $\lambda_{111}$ are shown in (c), and the corresponding $pp\rightarrow h_2\rightarrow h_1h_1$ rates in (d).   All results are presented for scenarios (black) S2, (blue) S3, and (red) S4. In (a,c) the 1$\sigma$ and 2$\sigma$ projected bounds on $\lambda_{111}$ at the HL-LHC~\cite{ATL-PHYS-PUB-2022-005} are shown in dark and light grey shaded regions, respectively.}
    \label{fig:lambdas}
\end{figure}

Figure~\ref{fig:lambdas} shows the same results as Fig.~\ref{fig:lambdas_MaxSM}, but for the future collider scenarios HL-LHC (S2), HL-LHC+FCCee (S3), and HL-LHC+ILC500 (S4). For comparison, the 1$\sigma$ and 2$\sigma$ projected bounds on $\lambda_{111}$ at the HL-LHC are also shown~\cite{ATL-PHYS-PUB-2022-005}.   Similarly to the current LHC results, when the $h_2\rightarrow h_1h_1$ rate is maximized, $\lambda_{111}$ can be up to a factor of two larger than the SM prediction, within the allowed projected 2$\sigma$ limit on $\lambda_{111}$ at the HL-LHC.  In the future lepton collider scenarios, the maximum allowed $\lambda_{111}$ is always within the 1$\sigma$ bounds at the HL-LHC.  This is due to the strong constraints on $|\sin\theta|$ and the perturbative unitarity bound on $a_2$ [see Eqs.~(\ref{eq:lam111}), (\ref{eq:pertunit})].  If the perturbative unitarity upper limit on $a_2$ is relaxed, the variation on $\lambda_{111}$ can be larger.

The maximum allowed $\lambda_{111}$ is up to a factor of three larger than the SM prediction in scenario S2.  This maximum occurs for $m_2\sim 1.5-3.5$~TeV, which may be difficult to discover through resonant projection of $h_2$ at the HL-LHC due to suppressed production rates at high masses.  

There are some interesting behaviors when $\lambda_{111}$ is maximized or minimized.  In particular, when $\lambda_{111}$ is maximized there are certain resonance masses for which $\lambda_{112}$ and the $pp\rightarrow h_2\rightarrow h_1h_1$ rate becomes nearly zero.  To understand these, we take the small mixing angle limits of the Higgs trilinear coupling and $\lambda_{112}$ in Eqs.~(\ref{eq:lam111}), (\ref{lambda211}):
\begin{eqnarray}
\lambda_{111}&=&\lambda_{111}^{\rm SM}\left[1+\left(a_2\frac{v_{\rm EW}^2}{m_1^2}-\frac{3}{2}\right)\sin^2\theta\right]+\mathcal{O}(\sin^3\theta),\label{eq:lam111_small}\\
\lambda_{112}&=&\left(2\,a_2v_{\rm EW}-\frac{2\,m_1^2+m_2^2}{v_{\rm EW}}\right)\sin\theta+2\,b_3\sin^2\theta+\mathcal{O}(\sin^3\theta),
\end{eqnarray}
where $\lambda_{111}^{\rm SM}$ is the SM prediction.  The expansion for the Higgs trilinear coupling $\lambda_{111}$ shows that it is maximized or minimized by choosing the largest allowed $|\sin\theta|$ and then maximizing or minimizing $a_2$.  From Eqs.~(\ref{eq:lam}), (\ref{eq:pertunit}), the numerical bounds on $a_2$ are
\begin{eqnarray}
-1.5+\mathcal{O}(\sin^2\theta)\leq a_2\leq 25.
\end{eqnarray}
Since the magnitude of the lower bound $a_2$ is much smaller than the upper bound, the minimum of $\lambda_{111}$ is near the SM value while the maximum can still have fairly large enhancements.  However, the perturbative unitarity upper bound on $a_2$ is not always saturated.  For smaller masses, global minimization places stronger bounds on $a_2$ than perturbative unitary~\cite{Chen:2014ask}.  As $m_2$ increases, the bounds on $a_2$ relax until they reach the perturbative unitarity bound.  Hence, the maximum $\lambda_{111}$ increases with $m_2$ and then flattens out when the allowed $a_2$ saturates $8\pi$ and $\sin\theta$ reaches the precision Higgs limits.  Then, in the multi-TeV range the narrow width constraints on $\sin\theta$ cause the maximum $\lambda_{111}$ to decrease.

Another interesting behavior is that when $\lambda_{111}$ is maximized, there are resonance masses at which $\lambda_{112}$ is very small.  From Eq.~(\ref{eq:lam111_small}) the maximum $\lambda_{111}$ has no dependence on $m_2$ except indirectly through $a_2$ and $\sin\theta$.  Once the model parameters are fixed, there is a mass for which $\lambda_{112}$ is small and nearly zero.
\begin{eqnarray}
m_2=\sqrt{2\left(a_2v_{EW}^2-m_1^2\right)}\left(1+\frac{1}{2}\frac{b_3v_{\rm EW}}{a_2v_{\rm EW}^2-m_1^2}\sin\theta\right)+\mathcal{O}(\sin^2\theta).
\end{eqnarray}
Assuming the perturbative unitarity constraint is saturated, then the numerical value of $m_2$ is
\begin{eqnarray}
m_2\approx 1.7~{\rm TeV}+0.03~{\rm TeV}\frac{b_3}{v_{\rm EW}}\sin\theta.
\end{eqnarray}
This is precisely where the very small $pp\rightarrow h_2\rightarrow h_1h_1$ rate occurs when maximizing $\lambda_{111}$ in the HL-LHC+FCCee and HL-LHC+ILC500 scenarios.  Hence, measurements of $\lambda_{111}$ and searches for the $h_2\rightarrow h_1h_1$ resonance can provide complementary information about the parameter space of the real singlet scalar model.

Higher order corrections to the $\lambda_{111}$ in the real singlet model can be $\mathcal{O}(50-100\%)$ depending on the scalar mass and coupling ranges~\cite{Curtin:2014jma,Chen:2017qcz,He:2016sqr, Kanemura:2016lkz}.  However, to account for these one-loop effects, a full one-loop electroweak correction to di-Higgs production in this model would need to be completed since the new scalar will contribute to more diagrams than just the trilinear Higgs production. While such a calculation will be necessary in interpreting any new physics signals, we leave that to later work and focus on the tree level results to gain an understanding of how much variation in the Higgs trilinear could still be viable in this model.

\section{Summary and conclusions}
\label{sec:conc}
In this paper we studied the di-Higgs rate in the simplest extension of the SM: the real singlet model with no additional symmetry.  This model is interesting in that it can generate resonant production of di-Higgs final states and change the Higgs trilinear coupling away from SM predictions.  We considered various scenarios for current and future colliders:
\begin{enumerate}[label=(\roman*)]
\item S1: Current LHC constraints from precision Higgs measurements and scalar searches~\cite{Lane:2024vur}.
\item S2: Projections for Higgs measurements and scalar searches at the HL-LHC~\cite{EuropeanStrategyforParticlePhysicsPreparatoryGroup:2019qin,Dawson:2022zbb}.
\item S3: Projections for Higgs measurements and scalar searches at the HL-LHC and a future circular $e^-e^+$ collider such as the CEPC or FCC-ee (HL-LHC+FCCee)~\cite{EuropeanStrategyforParticlePhysicsPreparatoryGroup:2019qin,Dawson:2022zbb}.
\item S4: Projections for Higgs measurements and scalar searches at the HL-LHC and a future linear $e^-e^+$ collider such as the 500 GeV ILC (HL-LHC+ILC500)~\cite{EuropeanStrategyforParticlePhysicsPreparatoryGroup:2019qin,Dawson:2022zbb}.
\end{enumerate}
Scenario S1 is an update on previous results in Ref.~\cite{Lewis:2017dme} for the HL-LHC with resonance masses up to 1 TeV, while scenarios S2-S4 go beyond and provide benchmarks for future high energy colliders for resonance masses up to 10 TeV.  

Considering these scenarios, the maximum allowed resonant di-Higgs $pp\rightarrow h_2\rightarrow h_1h_1$ rate  was found.  All theoretical constraints were considered, and we presented extensive discussions on the interplay between various constraints, the maximum resonant rate, and the corresponding scalar mixing angle and $h_2\rightarrow h_1h_1$ branching ratio. In scenarios S2-S4, the results are presented such that they are widely applicable for many future high energy collider possibilities.  For scenario S1, resonant di-Higgs production in this model can still be up to an order of magnitude larger than the SM di-Higgs rate.

We also studied how large deviations in the SM-like Higgs trilinear coupling $h_1-h_1-h_1$ can be in this model.  Here we also provided a discussion on the interplay of various model constraints, the behavior of the maximum and minimum of $\lambda_{111}$, and the resonant $pp\rightarrow h_2\rightarrow h_1h_1$ rates.  Boundedness and perturbative unitarity constraints force the minimum Higgs trilinear to be near the SM value while there can still be fairly large enhancements to $\lambda_{111}$.  Additionally, when $\lambda_{111}$ is maximized there is a value of the resonance mass for which the rate of $pp\rightarrow h_2\rightarrow h_1h_1$ is quite small, showing that resonant searches and measurements of $\lambda_{111}$ provide complementary information on the real singlet model parameter space.  We gave an analytical understanding of these behaviors.

The trilinear Higgs coupling can be up to a factor of 2.5 larger than the SM prediction in scenario S1 with resonance masses below 1 TeV, within the 2$\sigma$ projected bounds of the HL-LHC. In the real singlet model such large deviations in the trilinear coupling are consistent with a strong first order EW phase transition~\cite{Chung:2012vg,Curtin:2014jma,Huang:2016cjm,Barrow:2022gsu}, although additional study is needed to determine if these points can give rise to such a phase transition. In scenario S2 the trilinear Higgs couplings can be up a factor 3 larger than the SM prediction, just at the 2$\sigma$ sensitivity threshold for the HL-LHC.  In the future lepton collider scenarios, the allowed ranges of $\lambda_{111}$ are within 50$\%$ of the SM prediction and within the 1$\sigma$ HL-LHC constraints on $\lambda_{111}$. Of particular interest, the maximum value of $\lambda_{111}$ in the HL-LHC scenario (S2) is at or above the HL-LHC projected constraints on the Higgs trilinear coupling for new scalar mass in $1.5-3.5$ TeV range.  Such high masses may be difficult to discover through resonant production due to suppressed production rates.  Hence, measurements of the Higgs trilinear coupling through nonresonant di-Higgs production and resonant di-Higgs searches at the HL-LHC will provide sensitivity to complementary regions of parameter space.

\section*{Acknowledgments}
J.S., I.M.L., and M.A.S.A. are supported in part by the United States Department of Energy Grant No. DE-SC0017988.  M.S. is also supported by the United States Department of Energy under Grant Contract No. DE-SC0012704.  Digital data for
the plots and benchmark points have been uploaded with the arXiv version of this paper.

\appendix
\section{$h_2$ total width in the high mass region}
\label{app:GBET}
Here we give a short derivation of Eq.~(\ref{eq:GBET}).  In the high mass limit, $h_2\rightarrow W^\mp W^\pm$, $h_2\rightarrow ZZ$, and $h_2\rightarrow h_1h_1$ are the dominant decay channels of $h_2$ due to a longitudinal enhancement.  Hence the total width of $h_2$ can be approximated as
\begin{eqnarray}
\Gamma(h_2)\approx \Gamma(h_2\rightarrow W^\mp W^\pm)+\Gamma(h_2\rightarrow ZZ)+\Gamma(h_2\rightarrow h_1h_1).
\end{eqnarray}
The Goldstone Boson Equivalence Theorem relates these partial widths.
\begin{eqnarray}
\Gamma(h_2\rightarrow W^\mp W^\pm)\approx 2\,\Gamma(h_2\rightarrow ZZ)\approx 2\,\Gamma(h_2\rightarrow h_1h_1).
\end{eqnarray}
Hence, we have.
\begin{eqnarray}
\Gamma(h_2) \approx 2\Gamma(h_2\rightarrow W^\mp W^\pm)=2\,\sin^2\theta \Gamma_{\rm SM}(h_2\rightarrow W^\mp W^\pm),\label{eq:h2WidthGBET}
\end{eqnarray}
where we have applied the mixing angle suppression of Eq.~(\ref{eq:h2Rates}).

For the SM-like total width at $m_2$, we similarly have.
\begin{eqnarray}
\Gamma_{\rm SM}(h_2)&\approx& \Gamma_{\rm SM}(h_2\rightarrow W^\mp W^\pm)+\Gamma_{\rm SM}(h_2\rightarrow ZZ)\nonumber\\
&\approx& \frac{3}{2}\Gamma_{\rm SM}(h_2\rightarrow W^\mp W^\pm),\label{eq:SMh2WidthGBET}
\end{eqnarray}
where in the second line the Goldstone Boson Equivalence Theorem was applied.  

Hence, comparing Eqs.~(\ref{eq:h2WidthGBET}) and~(\ref{eq:SMh2WidthGBET}) as well as applying the high mass form of $\Gamma_{\rm SM}(h_2\rightarrow W^\mp W^\pm)$, we have Eq.~(\ref{eq:GBET}).
\begin{eqnarray}
\Gamma(h_2)\approx \frac{4}{3}\sin^2\theta\,\Gamma_{SM}(h_2)\approx \frac{m_2^3}{8\,\pi v_{\rm EW}^2}\sin^2\theta.
\end{eqnarray}

\section{Minima of Scalar Potential}
\label{app:min}
For completeness, we repeat the results of Refs.~\cite{Chen:2014ask,Lewis:2017dme}.  For details of their derivations, please refer to those references. There are six minima of the scalar potential, $(v, x)$. By definition, one is $(v, x) = (v_{\rm EW}, 0)$. There are two $v \neq 0$ solutions given by $(v, x) = (v_\pm, x_\pm)$ where
\begin{align}
    x_\pm &= \frac{v_{\rm EW} (3 a_1 a_2 - 8 b_3 \lambda) \pm 8 \sqrt{\Delta}}{4 v_{\rm EW} (4 b_4 \lambda - a_2^2)}, \\
    v_\pm^2 &= v_{\rm EW}^2 - \frac{1}{2 \lambda} (a_1 x_\pm + a_2 x_\pm^2), \\
    \Delta &= \frac{v_{\rm EW}^2}{64} (8 b_3 \lambda - 3 a_1 a_2)^2 - \frac{m_1^2 m_2^2}{2} (4 b_4 \lambda - a_2^2).
\end{align}
And, lastly, there are three $v=0$ solutions $x_i^0$ where $i=1,2,3$:
\begin{align}
    x_1^0 &= \frac{(2b_3 - \kappa^{1/3})^2 - 12 b_2 b_4}{6 b_4 \kappa^{1/3}} + \frac{b_3}{3 b_4}, \\
    x_2^0 &= \frac{(2b_3 - e^{2 i \pi / 3}\kappa^{1/3})^2 - 12 b_2 b_4}{6 b_4 e^{2 i \pi / 3} \kappa^{1/3}} + \frac{b_3}{3 b_4}, \\
    x_3^0 &= \frac{(2b_3 - e^{4 i \pi / 3}\kappa^{1/3})^2 - 12 b_2 b_4}{6 b_4 e^{4 i \pi / 3} \kappa^{1/3}} + \frac{b_3}{3 b_4},
\end{align}
with
\begin{align}
    \kappa &= -4 b_3(2 b_3^2 - 9 b_2 b_4) + 27 a_1 b_4^2 v_{\rm EW}^2 + 3 b_4 \sqrt{3 \Delta^0}, \\
    \Delta^0 &= -16 b_2^2 (b_3^2 - 4 b_2 b_4) - 8 a_1 b_3 v_{\rm EW}^2 (2 b_3^2 - 9 b_2 b_4) + 27 a_1^2 b_4^2 v_{\rm EW}^4.
\end{align}
The heat maps in Fig.~\ref{fig:compareb4} correspond to the points where $(v, x) = (v_{\rm}, 0)$ is the global minimum.
\section{Quantitative understanding of the rate maximization curves}
\label{app:kinks}

\begin{figure}[!htbp]
    \centering
        \subfigure[]{\includegraphics[width=0.41\textwidth,clip]{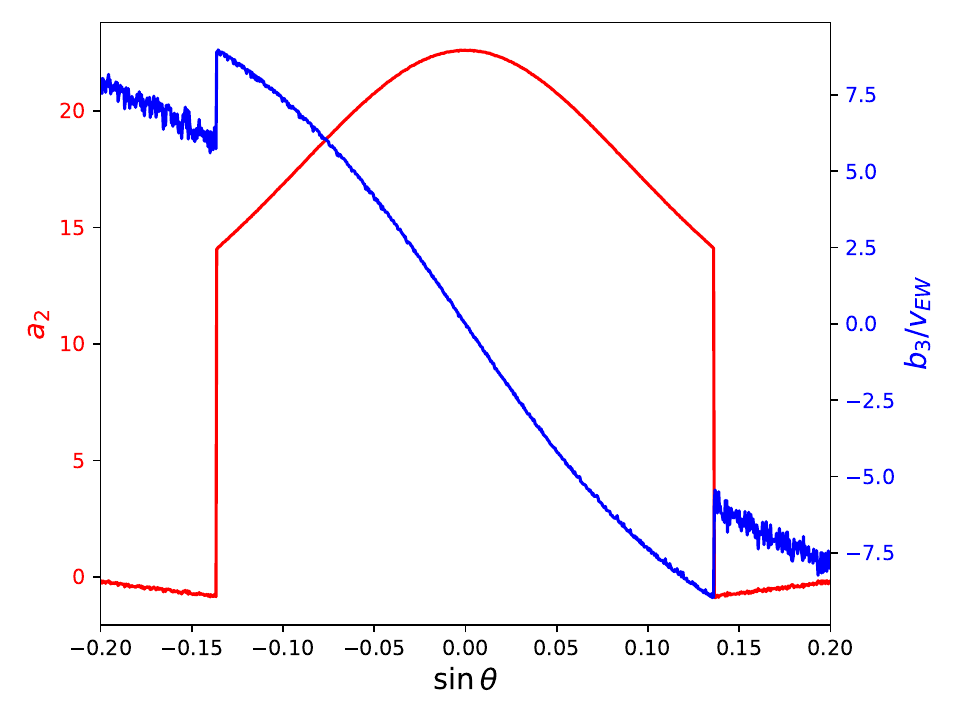}}
    \subfigure[]{\includegraphics[width=0.41\textwidth,clip]{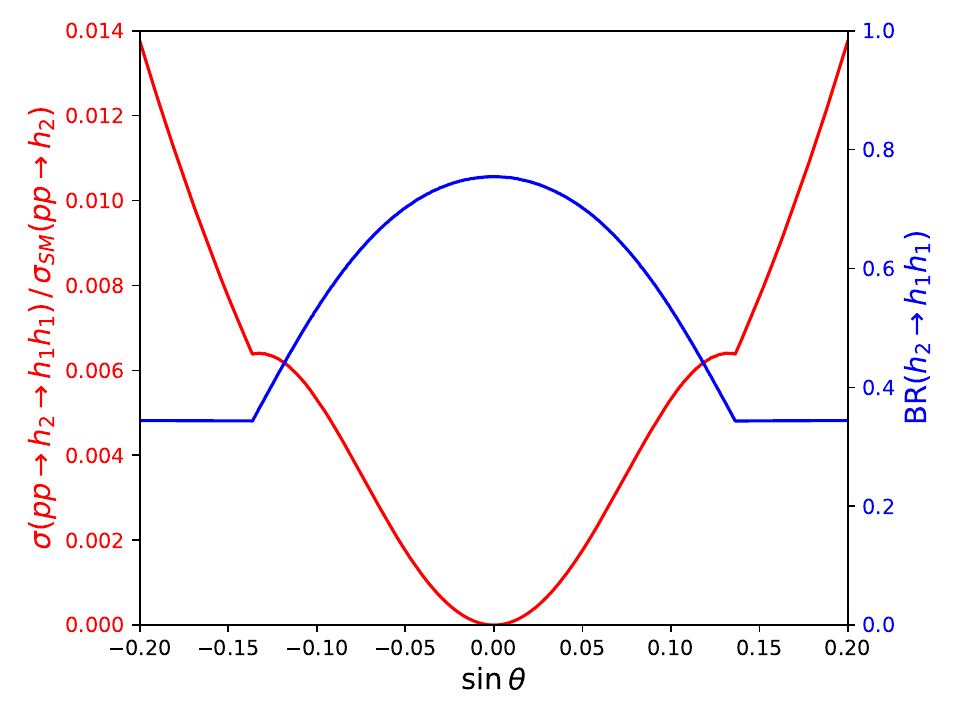}}\\[-2ex]
      \subfigure[]{\includegraphics[width=0.41\textwidth,clip]{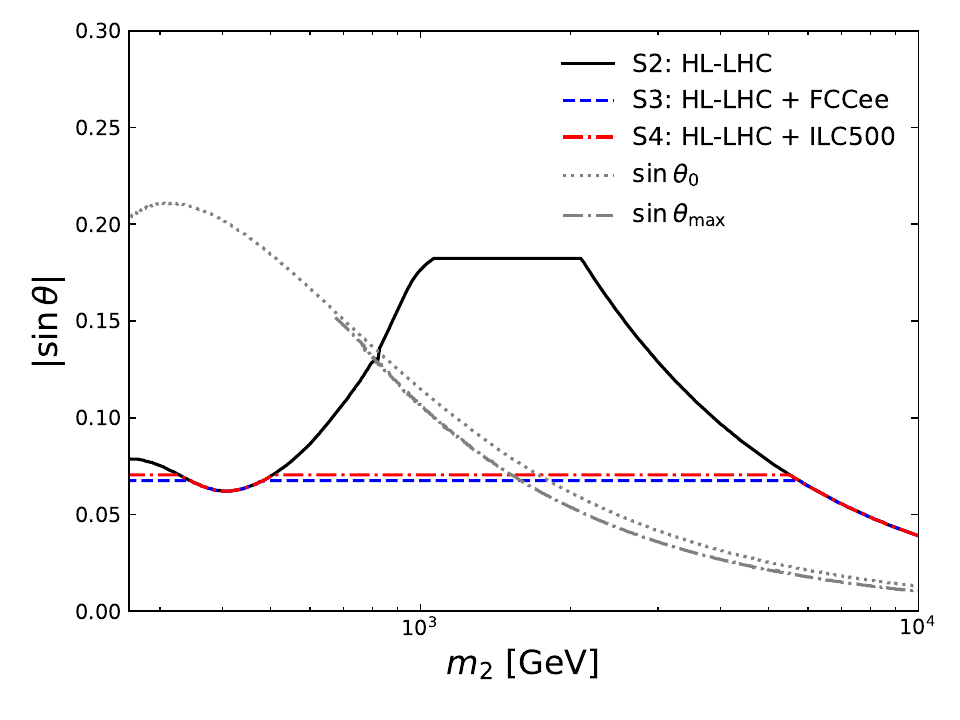}}
   \subfigure[]{\includegraphics[width=0.41\textwidth,clip]{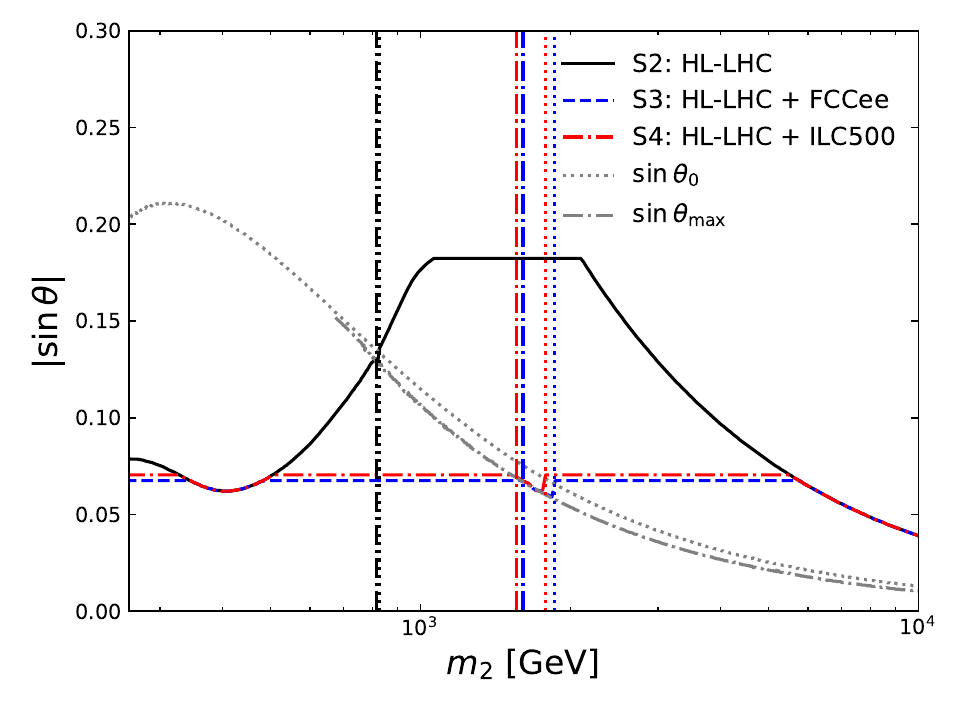}}\\[-2ex]
      \subfigure[]{\includegraphics[width=0.41\textwidth,clip]{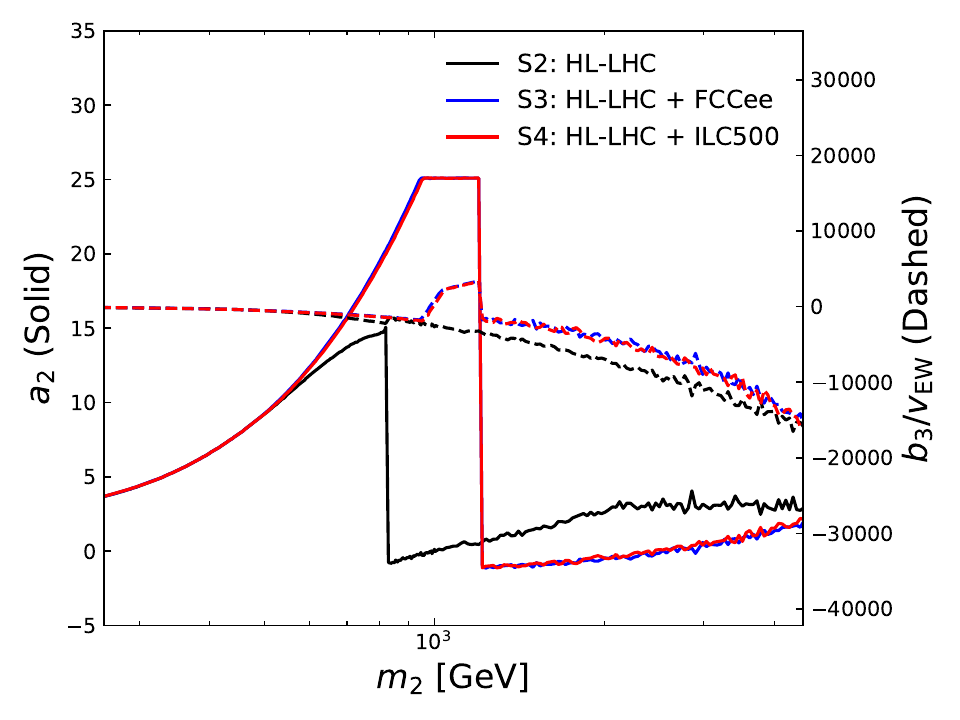}}
   \subfigure[]{\includegraphics[width=0.41\textwidth,clip]{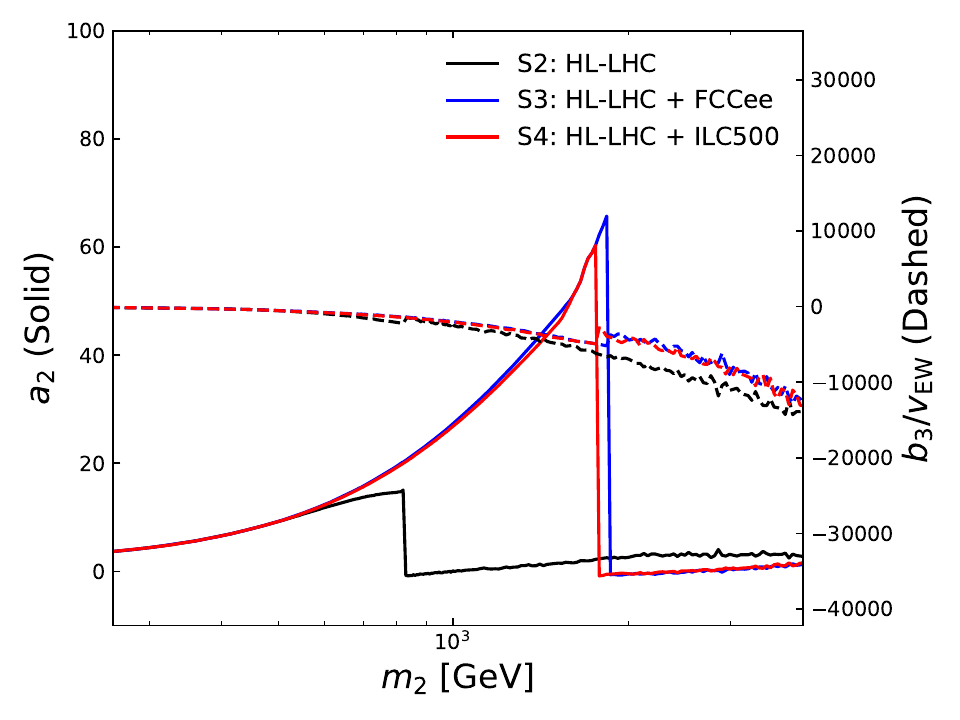}}\\[-2ex]
    \caption{(a) Values for (red) $a_2$ and (blue) $b_3/v_{\rm EW}$ that maximize the double Higgs rate for $m_2=800~{\rm GeV}$ as a function of mixing angle.   (b) The (red) maximum double Higgs rate normalized to SM-like production cross section and (blue) the corresponding ${\rm BR}(h_2\rightarrow h_1h_1)$ as a function of $\sin\theta$. (c,d) The mixing angles corresponding to the maximum rate as well as (dotted) $\sin\theta_0$ and (dot-dash) $\sin\theta_{\rm max}$.  In (d) the vertical lines show values of $m_2$ when $\sin\theta_0$ and $\sin\theta_{\rm max}$ cross the projected bounds.   (e,f) The values of (solid) $a_2$ and (dashed) $b_3$ that correspond to the maximum rate for scenarios S2-S4. In (c,e) all constraints are applied and (d,f) the perturbative unitarity bound on $a_2$ is neglected. In (c,d,e,f) scenario S2 is black, S3 blue, and S4 red.}
    \label{fig:A2_B3_comparison}
\end{figure}

Here, we will provide (numerical) explanations for the behaviors in the rate maximization curves and the associated branching ratios and $|\sin\theta|$ described in Sec.~\ref{sec:Results}. We will focus on the HL-LHC scenario first, although the explanations for HL-ILC+ILC500 and HL-LHC+FCCee are similar.  
From Fig.~\ref{fig:compareb4}, the rate is maximized on the borders of the allowed $b_3-a_2$ parameter plane. Which border of this plane (the left-hand side with smaller negative $a_2$ or right-hand side with larger positive $a_2$) that the maximum rate occurs on depends on the precise mass $m_2$ and scalar mixing angle.  Transitioning from the right border to the left border of the $b_3-a_2$ parameter plane in Fig.~\ref{fig:compareb4} changes the behavior of ${\rm BR}(h_2\rightarrow h_1h_1)$ when the di-Higgs rate is maximized.  To see this more clearly, Fig.~\ref{fig:A2_B3_comparison}(a) shows the $a_2,b_3$ values corresponding to the maximum rate for $m_2=800$~GeV as a function of $\sin\theta$.  The change from the left side to right side of the border of the $b_3-a_2$ parameter plane, as shown in Fig.~\ref{fig:compareb4}, occurs at $|\sin\theta|\sim 0.14$.  For $|\sin\theta|\gtrsim 0.14$, the parameter $a_2$ is negative and near zero indicating that the maximum rate is found on the left border of the allowed $b_3-a_2$ parameter plane while for $|\sin\theta|\lesssim 0.14$ the $a_2$ is positive and its magnitude is considerably larger, i.e. the maximum rate is on the right border of the $b_3-a_2$ parameter plane.  From Eq.~(\ref{lambda211}), when $a_2$ and $\sin\theta$ are small the $h_1-h_1-h_2$ coupling is largely independent of $a_2,b_3$ and ${\rm BR}(h_2\rightarrow h_1h_1)$ is largely independent of all model parameters. Explicitly, in the small mixing angle and large $m_2$ limit, the ${\rm BR}(h_2\rightarrow h_1h_1)$ can be expanded as
\begin{eqnarray}
{\rm BR}(h_2\rightarrow h_1h_1)&=&\frac{\left(1+2\,a_2\,v_{\rm EW}^2/m_2^2\right)^2}{3+\left(1+2\,a_2\,v_{\rm EW}^2/m_2^2\right)^2}+\mathcal{O}(\sin\theta,M_W^2/m_2^2,M_Z^2/m_2^2,m_1^2/m_2^2)\label{eq:BR_exp}\\
&=&\frac{1}{4}+\frac{3}{4}\frac{a_2\,v_{EW}^2}{m_2^2}+\mathcal{O}(\sin\theta,M_W^2/m_2^2,M_Z^2/m_2^2,m_1^2/m_2^2,a_2^3\,v_{\rm EW}^6/m_2^6)\nonumber
\end{eqnarray}
where in the second line we also expanded in the limit that $|a_2\,v_{EW}^2|\ll m_2^2$.  As is clear in the first line, the branching ratio can be large and have large $m_2$ dependence when $a_2$ is large.  Then, when $a_2$ is small, the branching ratio of $h_2\rightarrow h_1h_1$ is near $1/4$ (the Goldstone boson equivalence limit) and any residual dependence of the branching ratio on model parameters ($m_2,\,\sin\theta,\,a_2,\,b_3$) is small. Hence, this explains why the branching ratio shown in Fig.~\ref{fig:A2_B3_comparison}(b) is flat for $|\sin\theta|\gtrsim 0.14$.  Additionally, the maximum rate does not monotonically increase as $|\sin\theta|$ increases. Indeed, the rate has a local maximum right below $|\sin\theta|\approx0.14$ and then for $|\sin\theta|\gtrsim 0.14$ the rate monotonically increases again, as shown in Fig.~\ref{fig:A2_B3_comparison}(b). 

To quantify the discussion above, for each mass $m_2$ there is a $|\sin\theta\,|$ ($|\sin\theta_0|$) at which the maximal rates transition between large $a_2$ magnitudes to small $a_2$ magnitudes on opposite borders of the $b_3-a_2$ parameter plane.    As illustrated in Figs.~\ref{fig:A2_B3_comparison}(a,b) and Eq.~(\ref{eq:BR_exp}), for $|\sin\theta|\gtrsim |\sin\theta_0|$ the branching ratio of $h_2\rightarrow h_1h_1$ at the di-Higgs maximal rate is flat due to the magnitude of $a_2$ being smaller than for $|\sin\theta|\lesssim |\sin\theta_0|$.  Also, it can be seen in Fig.~\ref{fig:A2_B3_comparison}(b) that at $|\sin\theta_0|$ there is a sharp kink between decreasing and increasing maximal resonant di-Higgs rates.  That is, $|\sin\theta_0|$ is a local minimum of the di-Higgs resonance rate. Hence, there is a local maximum in the rate at a value of $|\sin\theta|$ ($|\sin\theta_{\rm max}|)$ near but below $|\sin\theta_0|$. As discussed above, at $m_2=800$~GeV $|\sin\theta_0|\approx0.14$.  These effects are relevant when the experimental limits on $|\sin\theta|$ and $|\sin\theta_0|$ coincide, as long as the corresponding values of $a_2,b_3$ are allowed within other constraints on the model, as we will discuss below.  In Fig.~\ref{fig:A2_B3_comparison}(c), we show the $|\sin\theta|$ corresponding to the maximum rate in scenarios S2-S4 as well as (dotted) $|\sin\theta_0|$ and (dash-dot) $|\sin\theta_{\rm max}|$. For the HL-LHC scenario there is a discontinuity in $|\sin\theta|$ at $m_2\sim 800$~GeV.  
For scenario S2, as $m_2$ increases the mixing angle corresponding to the maximum rate follows the projected limits in Fig.~\ref{fig:Max_sin}(b) until they cross the corresponding value of $|\sin\theta_{\rm max}|$ at $m_2\sim 800$~GeV. From there, the mixing angle at maximum rate follows  $|\sin\theta_{\rm max}|$ until it then crosses $|\sin\theta_0|$ at which point the value of $|\sin\theta|$ goes back to the projected limits.

As seen in Fig.~\ref{fig:A2_B3_comparison}(c), for the lepton collider scenarios S3 and S4 the $|\sin\theta|$ at maximal di-Higgs resonance rate do not have kinks at $|\sin\theta_0|$ or $|\sin\theta_{\rm max}|$.  These kinks would have been at $m_2\sim 2$~TeV.  However, in these cases and at these masses the corresponding $|a_2|$ values at maximal rates are far beyond the perturbative unitarity limit in Eq.~(\ref{eq:pertunit}).  
To make the importance of the perturbative unitarity limit clearer, in Fig.~\ref{fig:A2_B3_comparison}(d) we show the value of $|\sin\theta|$ corresponding to the maximum rates when the $a_2$ perturbative unitarity bound is neglected. As is clear, when the projected limits on $|\sin\theta|$ cross the $|\sin\theta_{\rm max}|$ curve, the mixing angle that maximizes the rate is $|\sin\theta_{\rm max}|$ as expected. For scenarios S3 and S4, this occurs in a very narrow band around $m_2\sim 2$~TeV between the vertical lines in Fig.~\ref{fig:A2_B3_comparison}(d), as mentioned in the previously.  Once the projected limits cross the $|\sin\theta_0|$ curve, the mixing angle that maximizes the rate reverts back to the limits in Fig.~\ref{fig:Max_sin}(b).

To further clarify the behavior of the mixing angle and branching ratios when maximizing the rates, in Fig.~\ref{fig:A2_B3_comparison}(e,f) we show the values of $a_2,b_3$ at the maximal resonant di-Higgs rate when (e) all constraints are considered and (f) when the perturbative unitarity limit on $a_2$ is relaxed. In either case, in the HL-LHC scenario at $m_2\sim 800$ GeV there is a transition from larger to smaller magnitudes of $a_2$, which behavior is discussed above.  For the HL-LHC+FCCee and HL-LHC+ILC500 scenarios, the perturbative unitary limit on $a_2$ is saturated at $m_2\sim 900$~GeV and then there is a transition between larger and smaller magnitude of $a_2$ at $m_2\sim 1.2$~TeV.  This is precisely where the branching ratio of $h_2\rightarrow h_1h_1$ flattens in Fig.~\ref{fig:BR}(b), since for small $a_2$ the branching ratio is relatively independent of model parameters, as shown in Eq.~(\ref{eq:BR_exp}). The transition from large to small $a_2$ occurs before the projected limits on $\sin\theta$ approach $|\sin\theta_{\rm max}|$ or $|\sin\theta_0|$. Hence, $|\sin\theta|$ never gets caught in the local rate maximum at $|\sin\theta_{\rm max}|$ and the discontinuities in the $|\sin\theta|$ that maximize the di-Higgs rate do not occur in scenarios S3 and S4.  If the perturbative unitarity bound on $a_2$ is neglected, $a_2$ is allowed to increase until $m_2\sim 1.9$~TeV where the projected limits on the mixing angle and $\sin\theta_0,\sin\theta_{\rm max}$ cross. This is precisely the discontinuity in scenarios S2 and S3 in Fig.~\ref{fig:A2_B3_comparison}(d), as we have mentioned earlier.  Then for $m_2\gtrsim 1.9$~TeV the magnitude of $a_2$ is much smaller.  However, with the perturbative unitarity limits that transition occurs much earlier and we get the results of Figs.~\ref{fig:BR} and~\ref{fig:A2_B3_comparison}(c).

In all scenarios, as $m_2$ increases there is a transition in the maximal resonance rate from larger to small $a_2$ magnitudes.  When $a_2$ is small, the $h_2\rightarrow h_1h_1$ branching ratio is relatively insensitive to model parameters, as shown in Figs.~\ref{fig:A2_B3_comparison}(a,b) and Eq.~(\ref{eq:BR_exp}). This explains why in Fig.~\ref{fig:BR}(b) that with increasing $m_2$ the branching ratio corresponding to the maximal rate at the HL-LHC transitions from relatively large values to near $1/4$ and being relatively flat as a function of $m_2$.  As discussed above in detail, for the HL-LHC scenario this occurs at $m_2\sim 800$~GeV and for the HL-LHC+FCCee and HL-LHC+ILC500 scenarios this occurs at $m_2\sim 1.2$~TeV.

\newpage 

\bibliographystyle{myutphys} 
\bibliography{example}

\end{document}